\documentclass[final,3p,times,twocolumn]{elsarticle}
\pdfoutput=1

\usepackage{graphicx}
\usepackage{amssymb}
\usepackage{natbib}
\usepackage{lineno}
\usepackage{subcaption}
\usepackage{graphicx}
\usepackage{tikz}
\usepackage{verbatim}
\usetikzlibrary{arrows}
\usepackage[nointegrals]{wasysym}
\usepackage{subfiles}
\usepackage{multirow}
\usepackage{placeins}
\usepackage{mathtools, cuted}
\usepackage{lipsum, color}
\usepackage{amsmath}
\usepackage{hyperref}
\usepackage{comment}
\usepackage{setspace}
\usepackage[colorinlistoftodos]{todonotes}
\definecolor{pds}{rgb}{1.0,0.5,0.5}
\definecolor{ntb}{rgb}{0.5,0.5,1.0}
\definecolor{Matt}{rgb}{0.5,1.0,0.5}

\newcommand{\K}{{$^{40}$K}}
\newcommand{\Mn}{{$^{54}$Mn}}
\newcommand{\Zn}{{$^{65}$Zn}}
\newcommand{\Ca}{{$^{40}$Ca}}
\newcommand{\Ar}{{$^{40}$Ar}}
\newcommand{\Y}{{$^{88}$Y}}
\newcommand{\Fe}{{$^{55}$Fe}}

\newcommand{\Tl}{$^{208}$Tl}
\newcommand{\Cs}{$^{137}$Cs}

\newcommand{\I}{$^{128}I$}
\newcommand{\Arthreenine}{{$^{39}$Ar}}

\newcommand{\ECStar}{EC$^{*}$}
\newcommand{\BREC}{$I_{EC}$}
\newcommand{\BRECStar}{$I_{EC^{*}}$}
\newcommand{\KSI}{KSr$_2$I$_5$:Eu}
\newcommand{\pplus}{p$^+$}
\newcommand{\Kalpha}{K$_{\alpha}$}
\newcommand{\Kbeta}{K$_{\beta}$}
\newcommand{\BMTAS}{$B_{M}$}
\newcommand{\BSDD}{$B_{S}$}

\newcommand{\keV}[1]{#1~keV}

\journal{Nuclear Instrument and Methods in Physics Research, A, }

\begin{document}

\doublespacing

\begin{frontmatter}

\title{\texorpdfstring{A novel experimental system for the KDK measurement of the $^{40}$K decay scheme relevant for rare event searches}{A novel experimental system for the KDK measurement of the K-40 decay scheme relevant for rare event searches} \\}

\author[label4]{M.~Stukel}
\author[label1,label2,label3]{B.C.~Rasco}
\author[label1,label2,label3]{N.T.~Brewer}
\author[label4]{P.C.F.~Di~Stefano \corref{cor}}
\ead{distefan@queensu.ca}
%\cortext[cor]{Corresponding author. Address: 64 Bader Lane, Kingston, Ontario, K7L 3N6, Canada}
\cortext[cor]{Corresponding author, Department of Physics, Engineering Physics and Astronomy, Queen’s University, 64 Bader Lane, Kingston, Ontario, K7L 3N6, Canada}

\author[label1]{K.P.~Rykaczewski}
\author[label7,label9]{H.~Davis}
\author[label7,label9]{E.D.~Lukosi}
\author[label4]{L.~Hariasz}
\author[label13]{M.~Constable}
\author[label14]{P.~Davis}
\author[label4]{K.~Dering}
\author[label5]{A.~Fija{\l}kowska}
\author[label6]{Z.~Gai}
\author[label7]{K.C.~Goetz}
\author[label1,label2,label3]{R.K.~Grzywacz}
\author[labelYYY,labelXXX]{J.~Kostensalo}
\author[label8]{J.~Ninkovic}
\author[label8]{P.~Lechner}
\author[label1]{Y.~Liu}
\author[label10]{M.~Mancuso}
\author[label11]{C.L.~Melcher}
\author[label10]{F.~Petricca}
\author[label6]{C.~Rouleau}
\author[label4]{P.~Squillari}
\author[label11]{L.~Stand}
\author[label1]{D.W.~Stracener}
\author[labelXXX]{J.~Suhonen}
\author[label1,label3,label12]{M.~Woli{\'n}ska-Cichocka}
\author{I.~Yavin}

\address[label4]{Department of Physics, Engineering Physics and Astronomy, Queen’s University, Kingston, Ontario, Canada}
\address[label1]{Physics Division, Oak Ridge National Laboratory, Oak Ridge, Tennessee, USA}
\address[label2]{Department of Physics and Astronomy, University of Tennessee, Knoxville, Tennessee, USA}
\address[label3]{Joint Institute for Nuclear Physics and Application, Oak Ridge, Tennessee, USA}
\address[label5]{Faculty of Physics, University of Warsaw, Warsaw, Poland}
\address[label6]{Center for Nanophase Materials Sciences, Oak Ridge National Laboratory, Oak Ridge, Tennessee, USA}
\address[label7]{Department of Nuclear Engineering, University of Tennessee, Knoxville, Tennessee, USA}
\address[label8]{MPG Semiconductor Laboratory, Munich, Germany}
\address[label9]{Joint Institute for Advanced Materials,University of Tennessee, Knoxville, Tennessee, USA}
\address[label10]{Max-Planck-Institut f\"{u}r Physik, F\"{o}hringer Ring 6, M\"{u}nchen, Germany}
\address[label11]{Scintillation Materials Research Center, University of Tennessee, Knoxville, Tennessee, USA}
\address[label12]{Heavy Ion Laboratory, University of Warsaw, Warsaw, Poland}
\address[label13]{TRIUMF, Vancouver, Canada}
\address[label14]{University of Alberta, Edmonton, Canada}
\address[labelYYY]{Natural Resources Institute Finland, Jokioinen, Finland}
\address[labelXXX]{University of Jyvaskyla, Department of Physics, P. O. Box 35, FI-40014, Finland}

\begin{abstract}

 Potassium-40 (\K) is a long-lived, naturally occurring radioactive isotope. The decay products are prominent backgrounds for many rare event searches, including those involving NaI-based scintillators. \K\ also plays a role in geochronological dating techniques. The branching ratio of the electron capture directly to the ground state of $^{40}$Ar has never been measured, which can cause difficulty in interpreting certain results or can lead to lack of precision depending on the field and analysis technique. The KDK (Potassium (K) Decay (DK)) collaboration is measuring this decay. A composite method has a silicon drift detector with an enriched, thermally deposited \K\ source inside the Modular Total Absorption Spectrometer. This setup has been characterized in terms of energy calibration, gamma tagging efficiency, live time and false negatives and positives. A complementary, homogeneous, method is also discussed; it employs a \KSI\ scintillator as source and detector.
\end{abstract}


\end{frontmatter}

\textbf{Journal :} Nuclear Instruments and Methods in Physics Research, Section A:
Accelerators, Spectrometers, Detectors and Associated Equipment
\section{\label{sec:Introduction}Introduction}
Natural potassium consists of three isotopes, of which potassium-40 (\K) is the only one that is radioactive. \K\ has a natural abundance of 0.0117(1)$\%$~\cite{be_table_2010} and is a main contributor to the radioactivity of the human body~\cite{thorne2003background}. Trace amounts of \K\ can also be found in a variety of minerals which, due to its very long half-life ($\sim$10$^{9}$ years), makes it widely used  for geochronological dating~\cite{aldrich_argon-40_1948,merrihue_potassium-argon_1966}.

\begin{figure}[ht]
  \centering
  \includegraphics[width=1.0\linewidth]{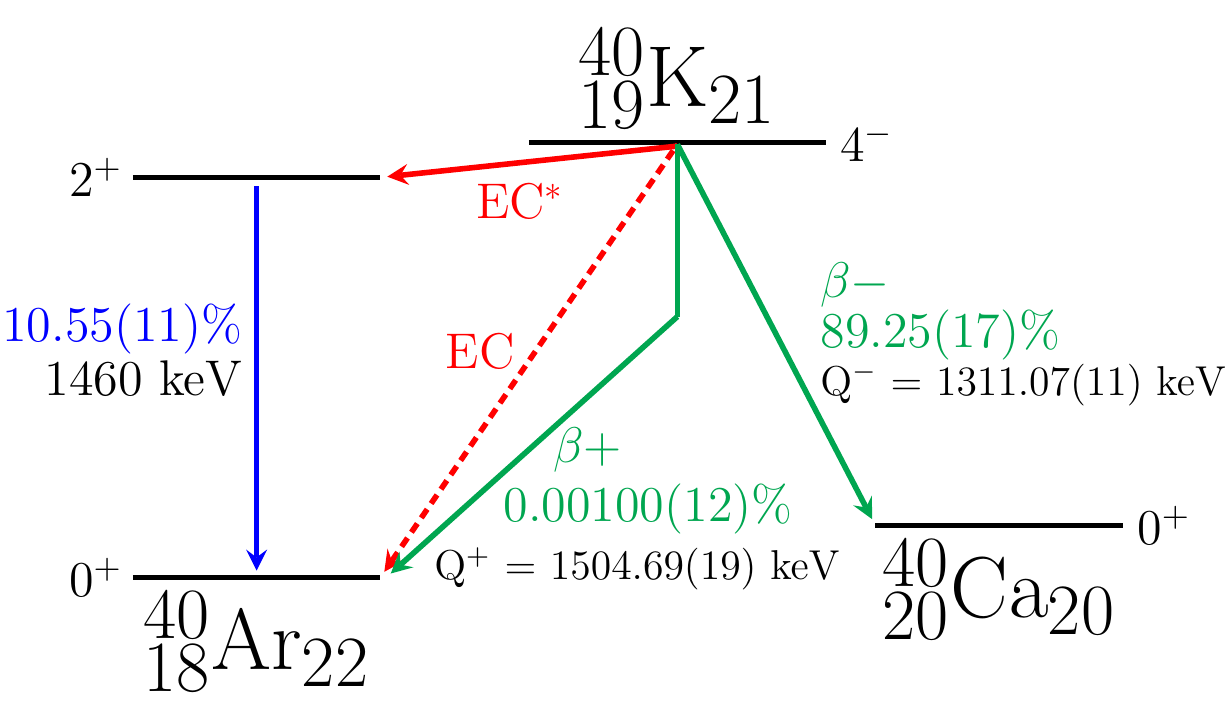}
	\caption{\label{Fig:K_40_decay_scheme}The decay scheme of $^{40}$K~\cite{be_table_2010}.  }
\end{figure}

The decay scheme of \K~can be seen in Fig.~\ref{Fig:K_40_decay_scheme}. The most recent and complete evaluations of the nuclear data for this isotope were performed in~\cite{be_table_2010} and~\cite{mougeot_x_40k_2009}. As shown, the $\beta ^-$ decay to \Ca~is the primary mode of decay with a partial half-life of 1.407(7)$\times$10$^9$~years. There also exist two modes of electron capture. The dominant one, with a partial half-life of 11.90(11)$\times$10$^9$~years, is to the excited state of \Ar\ (\ECStar). The electron capture process potentially releases a cascade of X-rays or Auger electrons, predominantly around 3~keV, from the shells of the argon daughter. As the nucleus de-excites it releases an additional 1460~keV $\gamma$-ray, or with a much smaller probability, another quantum like a conversion electron. The electron capture directly to the ground state of \Ar\ (EC) has never been experimentally observed. It behaves exactly like the \ECStar\ decay but without the associated de-excitation.

The decay scheme of \K\ was built by three channel probability ratios and the assumption that the decay scheme is complete. The first ratio is \BRECStar/$I_{\beta-}$ = 0.1182(12)~\cite{mougeot_x_40k_2009}, which is determined from the ratio of experimental half-lives, where $I_x$ represents the branching ratio of the assigned channel. The second ratio is $I_{\beta+}$/$I_{\beta-}$ = 1.12(14)$\times 10^{-5}$, from a single experiment~\cite{engelkemeir_positron_1962}. The final ratio is  \BREC/$I_{\beta+}$ = 200(100), from an extrapolation of the first and second forbidden unique transitions determined by the LogFT~\cite{noauthor_logft_nodate,gove_log-f_1971} calculator. 
 
 However, recent theoretical calculations have produced a large range of \BREC/$I_{\beta+}$ values. The National Nuclear Data Center gives a theoretical prediction of \BREC/$I_{\beta+}$ = 45.2(11)~\cite{chen_nuclear_2017}. Using data from~\cite{bambynek_orbital_1977}, reference~\cite{pradler_unverified_2013} calculates \BREC/$I_{\beta+}$ = 190, but this analysis makes the approximation of K-shell only electron captures. By performing higher order corrections, \cite{mougeot_improved_2018} calculates \BREC/$I_{\beta+}$ = 215.0(31). The range of \BREC/$I_{\beta+}$ causes almost an order of magnitude variation from \BREC\ = 0.045(12)--0.22(4)$\%$ in the branching ratio of the electron capture directly to the ground state.

The lack of experimental data and the range of theoretical values for this decay channel can pose issues for many fields. In nuclear physics, the EC decay is a rare example of an experimentally observable unique third forbidden electron capture decay (\textit{J$^{~\pi}$}(\K)~=~4$^-$ $\rightarrow$~\textit{J$^{~\pi}$}(\Ar)~=~0$^+$). Knowledge of this branching ratio will inform and test nuclear models. In geochronology, the K/Ar and derivative \Ar/\Arthreenine~dating methods are used extensively in a wide range of environments and time periods. As discussed in~\cite{min2000test,begemann_call_2001,renne2010joint}, an outstanding issue of these dating methods is verifying the relevance (or existence) of the EC decay. Implications for K-Ar and Ar-Ar dating are reviewed in~\cite{carter2020production}.

Finally, for many rare-event searches, the presence of \K~provides a challenging radioactive background. Due to the chemical similarity between K and Na, trace amounts of \K~can be found in ultra-radiopure scintillating crystals grown from NaI powders. Experiments using such scintillators include ANAIS~\cite{amare_first_2019}, COSINE-100~\cite{mouton_initial_2018}, COSINUS~\cite{angloher2020cosinus},
DM-Ice17~\cite{de_souza_first_2017}, 
PICO-LON~\cite{fushimi_dark_2016},
SABRE~\cite{antonello2019sabre,antonello2019monte}, and DAMA/LIBRA~\cite{first_DAMMA_2018} (referred to as DAMA from here on).

Of the active NaI dark matter searches, only DAMA has claimed to observe a dark matter signal. The effect of the \K\ background on the interpretation of the DAMA claim has been investigated~\cite{pradler_unverified_2013,stukel_characterization_2018}. The EC decay presents an unknown background  directly in the 2--6~keV energy signal region which needs to be understood.

Given the above, there is clear motivation to measure the direct to ground state electron capture of \K. In this paper we present the detector of the KDK (potassium (K) decay (DK))~\cite{di_stefano_kdk_2017} experiment. This article is structured in the following way: Section~\ref{sec:Exp_Setup} will discuss the experimental method and a technical description of two detector methods (homogeneous and composite) used to obtain this measurement. Section~\ref{sec:Detec_Charac} will detail the integration and performance of the composite method detector, including data reduction and energy calibration. Section~\ref{subsec:Efficiency} gives $\gamma$-ray efficiencies, live time considerations and predicted sensitivity of the composite setup.

\section{\label{sec:Exp_Setup}The detector and experimental setup}

 The KDK experiment consists of an inner detector with an energy sensitivity of a few keV or better surrounded (close to 4$\pi$~sr coverage) by an outer detector with higher energy (30--53000~keV) sensitivity. Ideally, the particles emitted by the \K~electron capture process trigger the inner detector and open a coincidence window with the outer detector. If a $\gamma$-ray is detected during this period the event is classified as an \ECStar.  If not, the event is classified as an EC. In practice, various factors complicate the analysis, discussed further in~\ref{sec:FalseNegative}. By distinguishing between EC and \ECStar~events, the ratio $\rho = I_{EC}/I_{EC^{*}}$ can be determined and used in building the \K\ decay scheme. 

 Two different methods are studied for the inner detector. The composite method consists of a separate low energy detector and \K~source.  This setup is discussed in Sections~\ref{subsec:SDD},~\ref{sec:Detec_Charac} and~\ref{subsec:Efficiency}. The homogeneous method contains the \K~source in a bulk scintillator and is discussed in Section~\ref{sec:Detec_Charac_KSI}.  The outer detector, used for both methods, is described in Section~\ref{subsec:MTAS}. 
 
 These methods were built in parallel with the founding design principles being complementary. The homogeneous method benefits from the \K\ being uniformly distributed throughout the bulk of the scintillator. The X-rays and Auger electrons produced during the decay of \K\ have a near-zero probability of escaping due to the dimensions of the scintillator being orders of magnitude larger than their absorption length. Conversely the composite method has excellent energy resolution which will successfully distinguish between the different electron capture transitions (i.e. \Kalpha\ and \Kbeta) and allow us to investigate backgrounds. In addition, the energy threshold of the composite inner detector will be considerably lower than that of the bulk scintillator and the external source will allow for dedicated background and calibration runs. At the start of development it was unclear whether each or both of these methods would produce a successful $\rho = I_{EC}/I_{EC^{*}}$ measurement. As such, both ideas were pursued with the details presented below.

\subsection{\label{subsec:MTAS}Outer detector: Modular Total Absorption Spectrometer}

\begin{figure}[ht]
  \centering
  \includegraphics[width=0.9\linewidth]{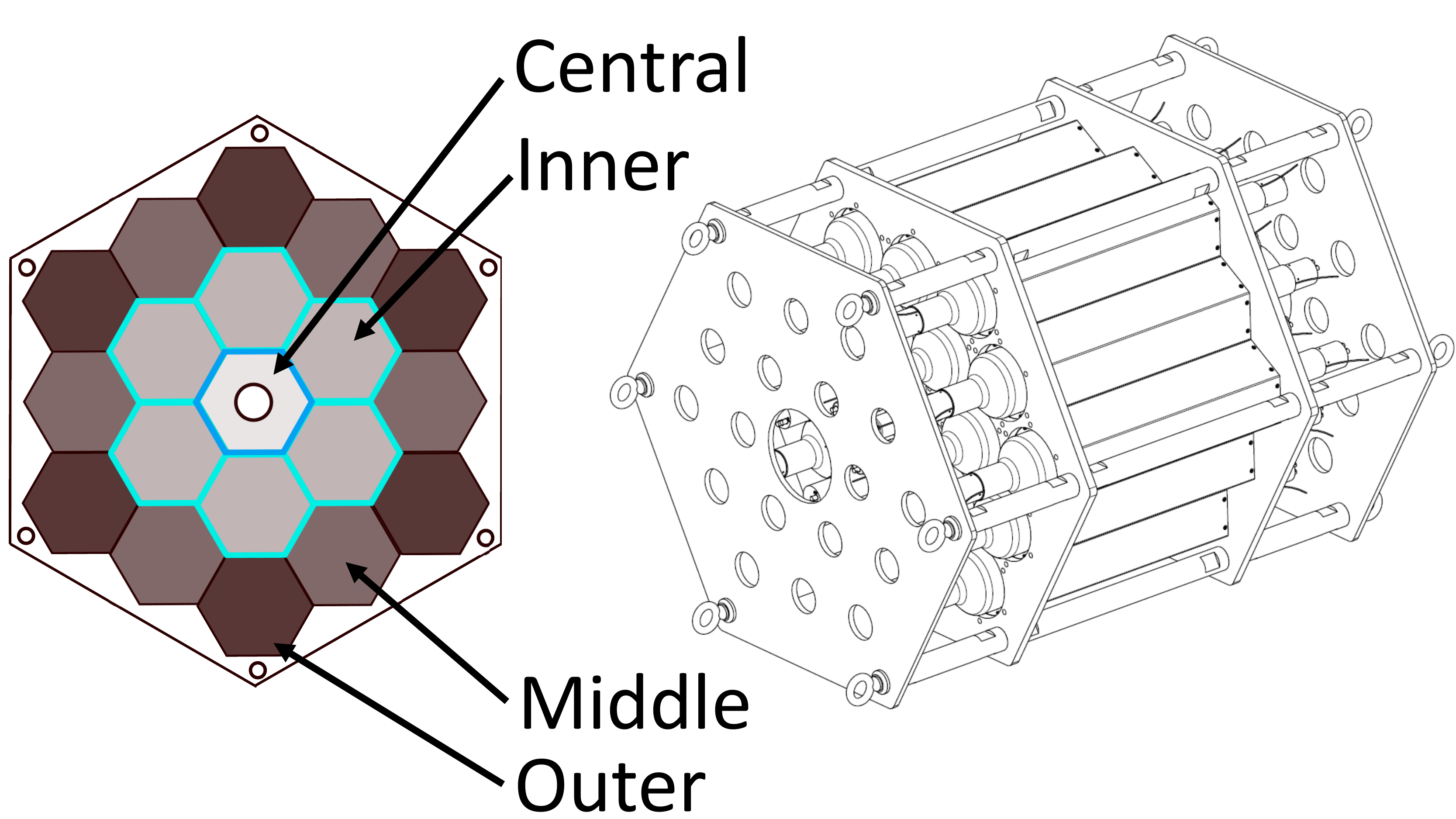}
	\caption{\label{Fig:MTAS_Schematic.png}Schematic of the Modular Total Absorption Spectrometer. The different layers (Central, Inner, Middle and Outer) are shown. Based on~\cite{wolinska-cichocka_modular_2014}.}
\end{figure}

The Modular Total Absorption Spectrometer (MTAS)~\cite{karny_modular_2016} at Oak Ridge National Laboratory (ORNL) is used as the outer detector for the KDK experiment. MTAS was created to study complex $\beta$ decays of reactor fission products~\cite{fijalkowska_impact_2017,rasco_decays_2016,rasco_complete_2017} and is composed of 19 NaI(Tl) hexagonal-shaped scintillators (53.34~cm in length $\times$ 17.6~cm across), with a total mass of close to a metric tonne. The centre module has a 63.5~mm-diameter through-hole, inside of which the inner detectors are located. The crystals are arranged in a honeycomb pattern with the different layers referred to as the central module, the inner ring, the middle ring, and the outer ring, as shown in Fig.~\ref{Fig:MTAS_Schematic.png}. These are classified by their distance from the centre of MTAS. Each crystal is enclosed by a carbon fiber housing, with thin layers of silicon putty, stainless steel, and Teflon. Two 12.7~cm-diameter ETI9390 Photo-Multipliers (PMTs), one on each end, are placed on every crystal in the Inner, Middle and Outer layers. The centre crystal has six 3.81~cm-diameter ETI9102 PMTs on each end. All PMTs were made by ET-Enterprises (Uxbridge, UK). In order to increase the efficiency with which MTAS can capture escaping gammas in the composite method, a 5.58~cm diameter $\times$ 25.4~cm length NaI(Tl) crystal plug from Saint-Gobain can be placed inside the central module. The crystal is surrounded by a 1.575~mm thick aluminum housing and has a ETI9266 PMT attached to the back. The total configuration (MTAS + plug) leaves a $\sim 1.4\%$ numerical aperture for escaping particles.  MTAS is surrounded by 2.5~cm of lead shielding plus additional layers of lead blankets. 

Signals from the PMTs are fed into Pixie-16 digital gamma finders from XIA LLC (Hayward, CA)~\cite{xia_llc_pixie-16_2009}. MTAS is connected to 49 Pixie-16 channels, one for each PMT, and each is triggered independently. These signals are digitized by a 12-bit analogue-to-digital converter at 100~MHz. The digital signal processing section of the Pixie-16 cards uses an optimized trapezoidal filtering algorithm that determines the energy (pulse height) and time-stamp of the signal~\cite{nathan_github}. 

Coincidence between different modules is enforced using a global rolling window. After one channel is triggered, with a selected coincidence window of $dt$, the algorithm checks if the subsequent triggered module time-stamp is within $dt$. If it is, the event is classified as \textit{in coincidence} and the window is extended for another $dt$. This is repeated until no more events are found within the extended coincidence window. Handling of pileup within the coincidence window is explored in Section~\ref{subsec:Dead Time Considerations}. The exact coincidence window can be varied in the offline analysis in order to understand the impact of different backgrounds. 

\begin{figure}[ht]
  \centering
  \includegraphics[width=1.0\linewidth]{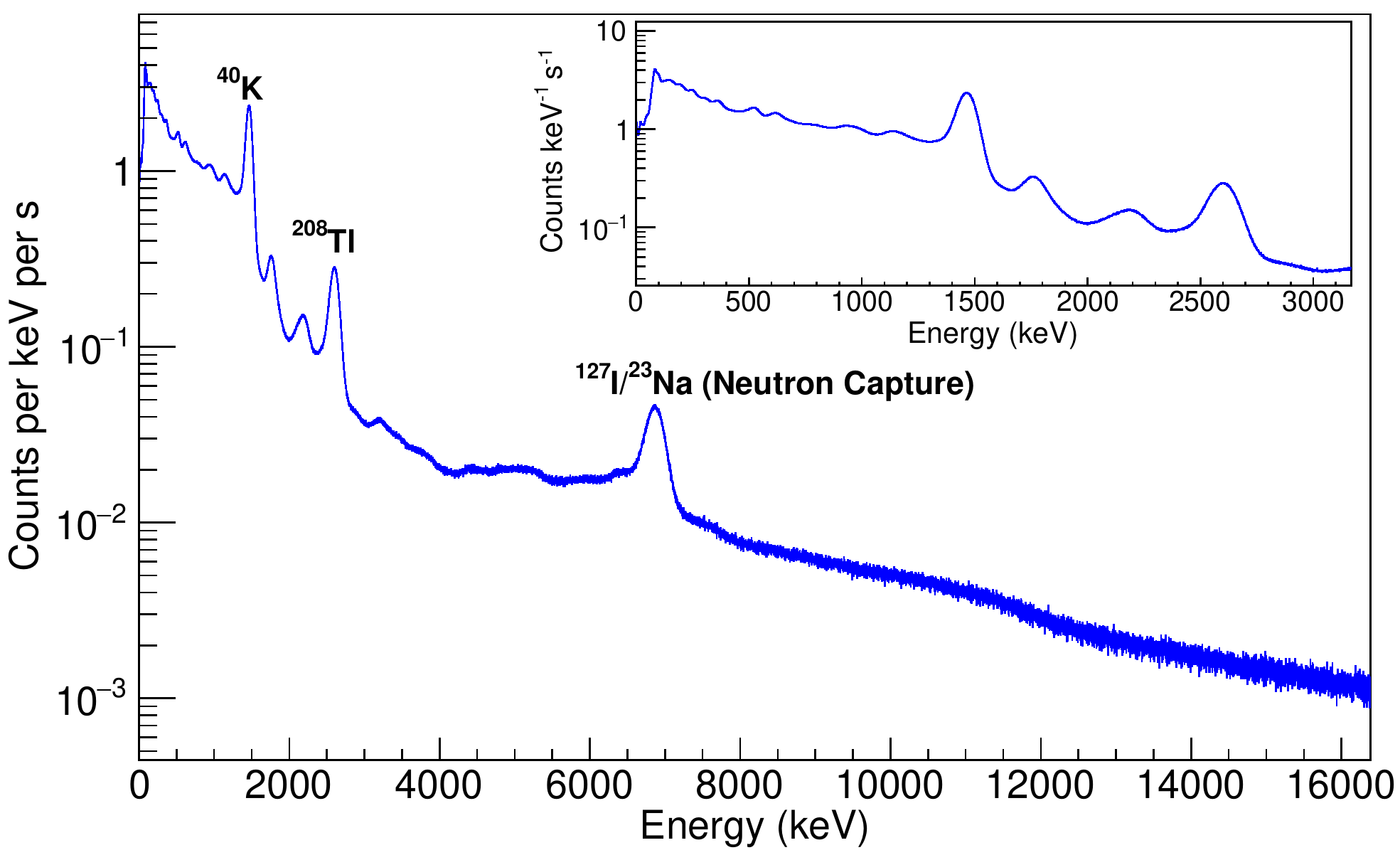}
	\caption{\label{Fig:MTAS_BKG_Energy_Spectrum.eps} The MTAS background energy spectrum, triggered by MTAS itself. The three largest background peaks visible are the \K: 1460~keV gamma peak, the \Tl: 2614~keV gamma peak, and the overlapping 6800~keV \textsuperscript{127}I and  \textsuperscript{23}Na neutron capture peaks. The insert shows a close-up of the spectrum in the 0 to 3000 keV range.}
\end{figure}

Like typical NaI(Tl) scintillators, those that make up MTAS contain a certain amount of \K~contamination. In addition, natural contamination from the surrounding environment and cosmogenic particles are easily detected by the massive device. The intrinsic background energy spectrum, when no external source is present, is shown in Fig.~\ref{Fig:MTAS_BKG_Energy_Spectrum.eps}. Thresholds of 30~keV for individual MTAS modules arise from the combined effects of multiple photomultipliers mounted symmetrically on each module and the NaI(Tl) signal processing. These thresholds have been experimentally verified~\cite{karny_modular_2016}. The energy resolution is 92~keV~full width at half maximum (FWHM) at the 1460~keV peak. The total count rate as a function of coincidence window is  given in Table~\ref{tab:MTAS_BKG_Rate}. This table shows an approximately two percent variation in count rate between the largest and smallest coincidence windows that were measured. The higher coincidence windows have a decreased rate due to composite background events forming (i.e. two background events are being added together). Performance of the MTAS detector is further discussed elsewhere~\cite{karny_modular_2016,fijalkowska_first_2014}. 

\begin{table}[ht]
\centering
\begin{tabular}{cc} 
    
\hline        
Coincidence Window ($\mu$s) & Rate (counts/s)  \\\hline
1.0 & 2652.56(15) \\
2.0 & 2639.51(15) \\
4.0 & 2618.92(15) \\\hline
\end{tabular}
\caption{Background rate in the MTAS detector as a function of coincidence window. Only events from 0 - 16000~keV are considered.}
\label{tab:MTAS_BKG_Rate}
\end{table}

\subsection{\label{subsec:SDD}Composite inner detector: large area silicon drift detector}

A large area silicon drift detector (SDD)\footnote{A large area avalanche photodiode (APD) was also tested as a candidate for the KDK experiment. The results of the APD testing can be seen in~\cite{stukel_characterization_2018}.} is used for the inner portion of the composite method. This type of detector  has been shown to successfully operate above liquid nitrogen temperatures while maintaining good energy resolution with greater than 90$\%$ quantum efficiency for 1-10 keV X-rays~\cite{lechner_silicon_1996,struder_silicon_1998}. Our detector was fabricated by the Halbleiterlabor of the Max-Planck-Society in Munich, Germany.

\subsubsection{\label{subsec:SDD_Operating_Principal}SDD operating principle}

The SDD functions on the principle of sideward depletion designed by Gatti and Rehak in 1983~\cite{gatti_semiconductor_1984}. In our configuration, a n-type silicon wafer is depleted by a small n$^+$ anode contact. Concentric \pplus\ electrode rings with increasing negative bias are placed around the anode. On the opposite end of the chip is a large \pplus\ planar cathode with a strong negative bias. In operating conditions, this setup creates an electric potential with the anode at the minimum. The potential energy distribution in the SDD can be seen in~\cite{struder_silicon_1998}. Any electrons produced inside the depleted region are transported to this minimum and are readout using a source-follower configuration~\cite{niculae_optimized_2006}. The holes that are generated are collected by the \pplus~implanted regions.

\begin{figure}[ht]
  \centering
  \includegraphics[width=1.0\linewidth]{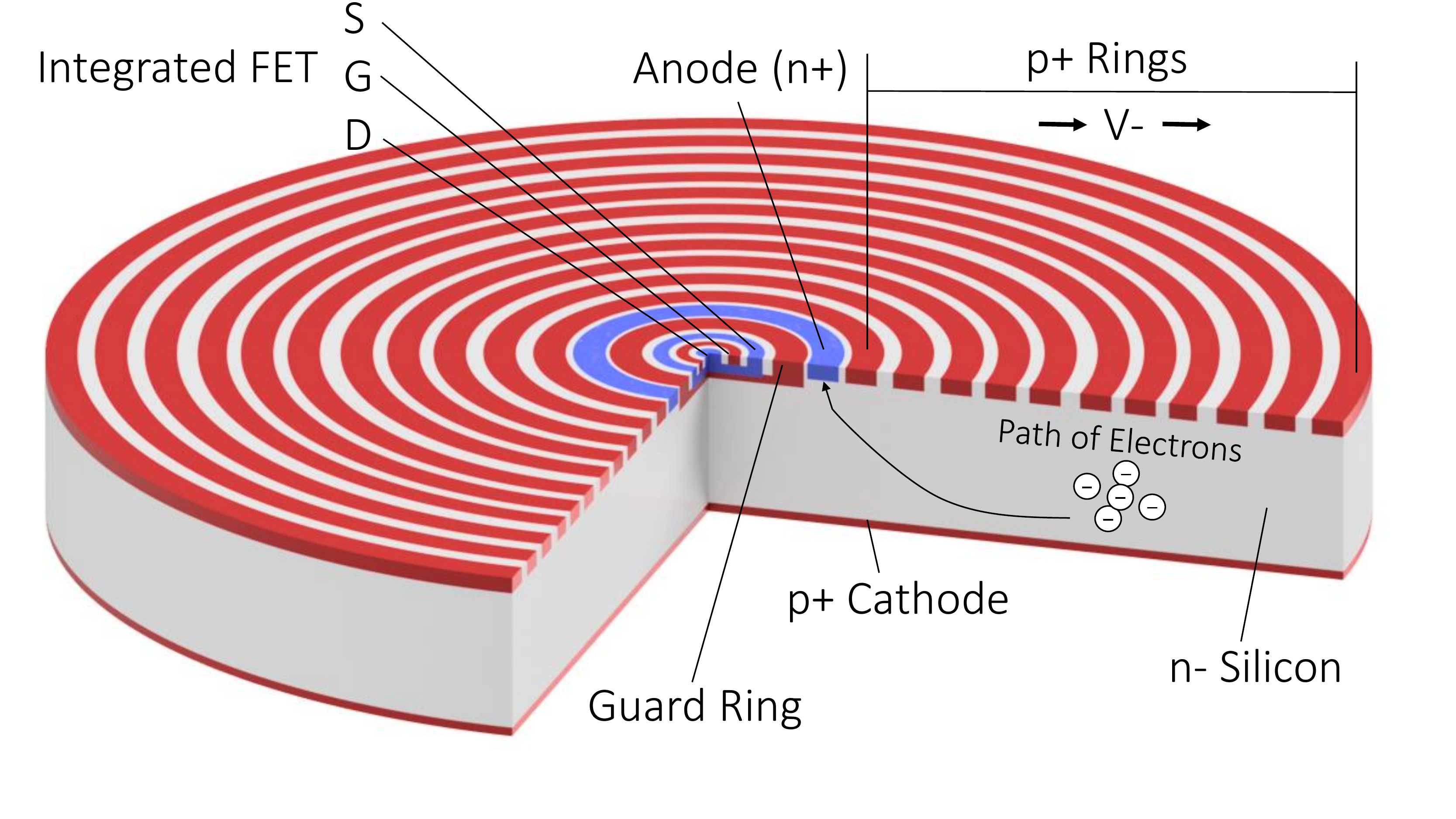}
	\caption{\label{Fig:SDD_Wafer_Schematic.pdf}Schematic tri-dimensional view of a SDD with integrated FET. Based on figure in~\cite{niculae_optimized_2006}.}
\end{figure}

Integrated onto the silicon chip is a field effect transistor (FET) whose gate is directly connected to the surrounding anode. The integrated FET avoids the use of bond wires thus minimizing the capacitance of the system. A guard ring structure is used to insulate the rest of the FET from the chip environment. A schematic for the SDD substrate can be seen in Fig.~\ref{Fig:SDD_Wafer_Schematic.pdf}. 

In order to avoid charge build-up in the anode, a self-reset mechanism is implemented. As the electrons arrive at the anode/gate connection it becomes more negatively charged. Eventually, the gate will reach a specific potential difference causing a weak avalanche region to form in the channel between the gate and the drain. The holes that are generated in this region are collected by the gate and compensate for the electrons by increasing its potential. Further details about the source-follower readout system and self-reset operation can be found in~\cite{fiorini_continuous_1999} and~\cite{fiorini_charge-sensitive_2002}.

\subsubsection{\label{subsec:SDD_Technical_Design}SDD implementation and integration into MTAS}

\begin{figure}
\centering
\begin{subfigure}[b]{0.8\linewidth}
   \includegraphics[width=1.0\linewidth]{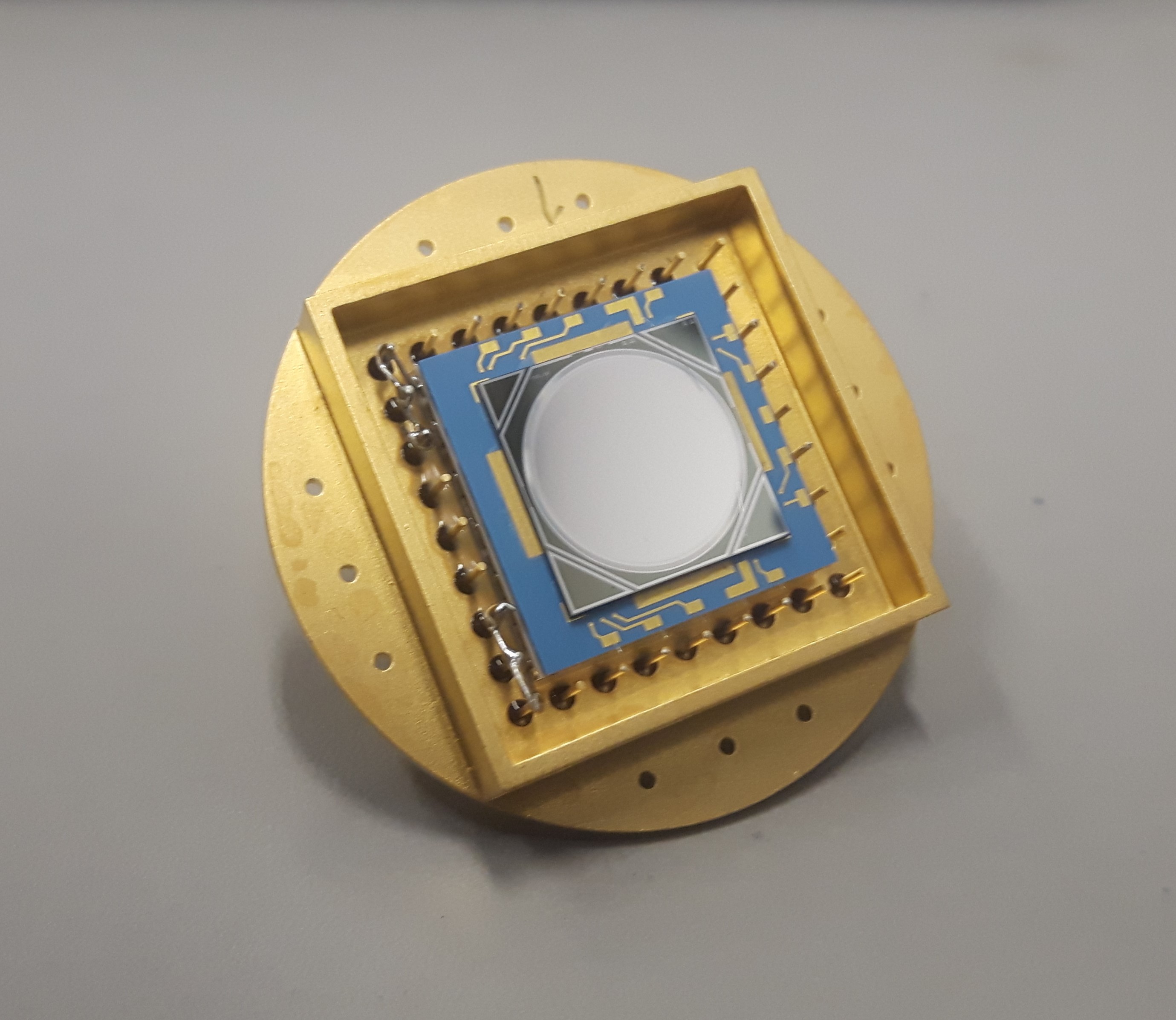}
   \caption{}
   
\end{subfigure}

\begin{subfigure}[b]{0.8\linewidth}
   \includegraphics[width=1.0\linewidth]{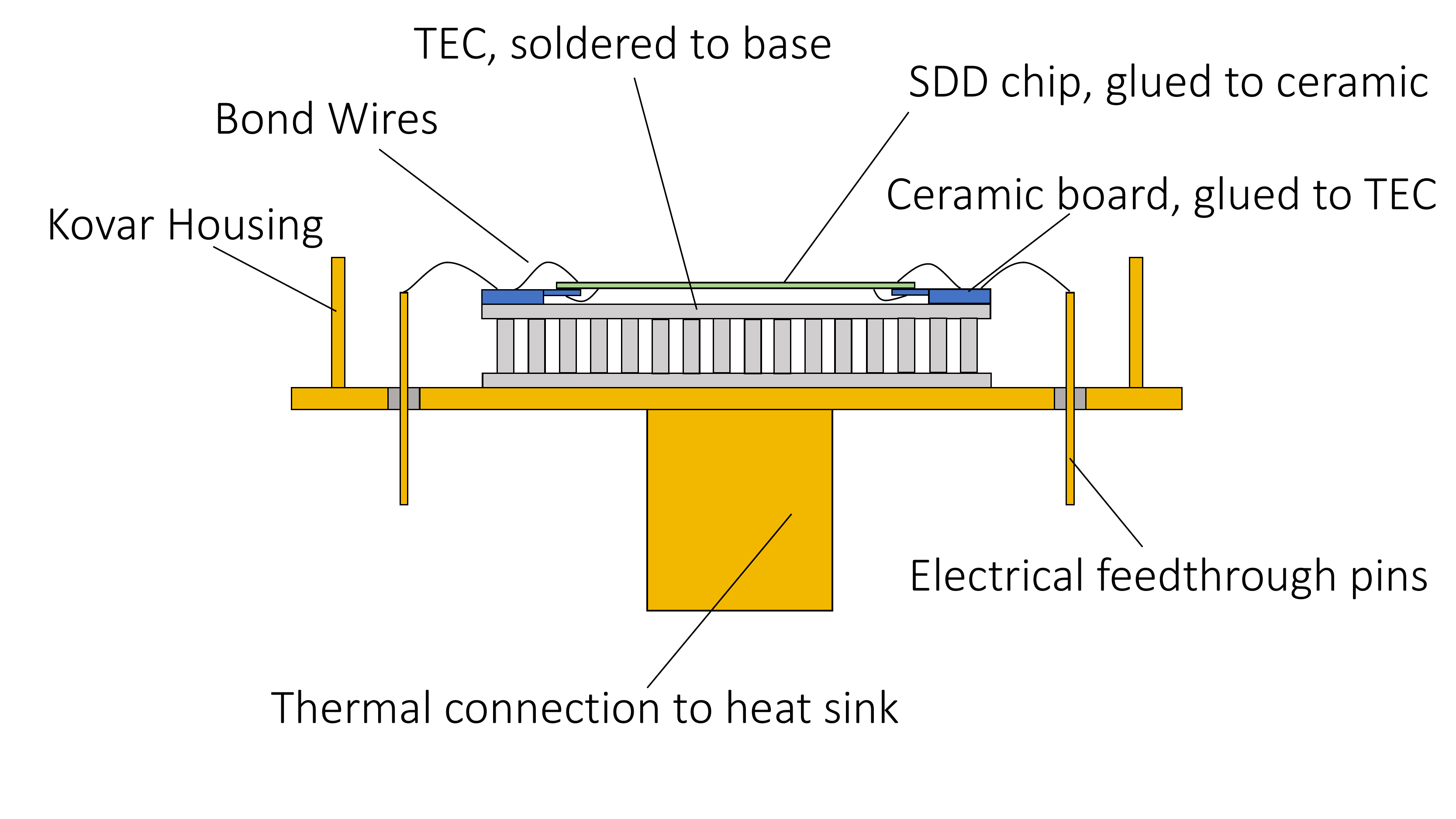}
   \caption{}
   
\end{subfigure}
\caption{\label{Fig:SDD_Real_w_Schematic}(a) SDD detector in the Kovar housing, front facing. (b) Schematic cross-section of the SDD assembly.}

\end{figure}

The SDD housing and vacuum setup is designed to minimize the amount of material within close proximity of the source. This is to reduce the scattering of the gammas in non-detecting areas. The SDD used in the KDK experiment is a cylindrical silicon volume with an active surface area of 100 mm$^2$. It is 450~$\mu$m thick with dead layers of silicon (\pplus, 30~nm), SiO$_2$ (20~nm) and aluminum (30~nm) on the radiation facing side. The SDD is glued to a printed (18 $\times$ 18 mm) ceramic board and connected together with aluminum bond wires. Another connection is made from the board to the electronics via feedthrough pins.  Underneath the board is a thermoelectric cooler (TEC), Model No. 1TML10-18$\times$18-15-000W from Thermion Company, Odessa, Ukraine. The TEC is 3.6 mm thick, made out of Bi$_2$Te$_3$ with 0.5 mm Al$_2$O$_3$ coating on both ends. The ensemble is housed in a custom designed Kovar unit that has 32 feedthrough pins and a thermal connection (M8 threading) to the heat sink. Fig.~\ref{Fig:SDD_Real_w_Schematic} shows a schematic of the SDD housing setup alongside a front visual of the actual detector.

The output signal of the SDD is connected to a spectroscopy pre-amplifier (Amptek-A275) residing on a printed circuit board (PCB) that is connected directly to the Kovar housing. The signal is then passed through a  Gaussian shaping amplifier (Ortec 672, residing outside the SDD vacuum system) and fed directly into the MTAS data acquisition system (DAQ).

To improve the signal-to-noise ratio of the SDD, it is cooled by means of an aluminum rod threaded onto the thermal sink of the Kovar housing. The rod is  clamped to a copper piece that thermally connects it to a copper cooling loop. The copper loop is a 4.76~mm inner diameter and 6.35~mm outer diameter tube that extends for 1.05~m to a ISO-100 liquid feedthrough flange. The cooling loop is powered by the RTE-140 refrigerated Bath/Circulator from Thermo Fisher Scientific, USA. A 50/50 mixture of filtered tap water and laboratory grade ethylene glycol is used as the cooling liquid. Temperatures of $-10^\circ$C can be reached at the SDD using this method. Additional cooling is provided by the TEC causing the SDD to reach the operating temperature of $-20^\circ$C. The temperature is measured using an integrated diode that resides on the SDD.  

\begin{figure*}[ht]
  \centering
  \includegraphics[width=1.0\linewidth]{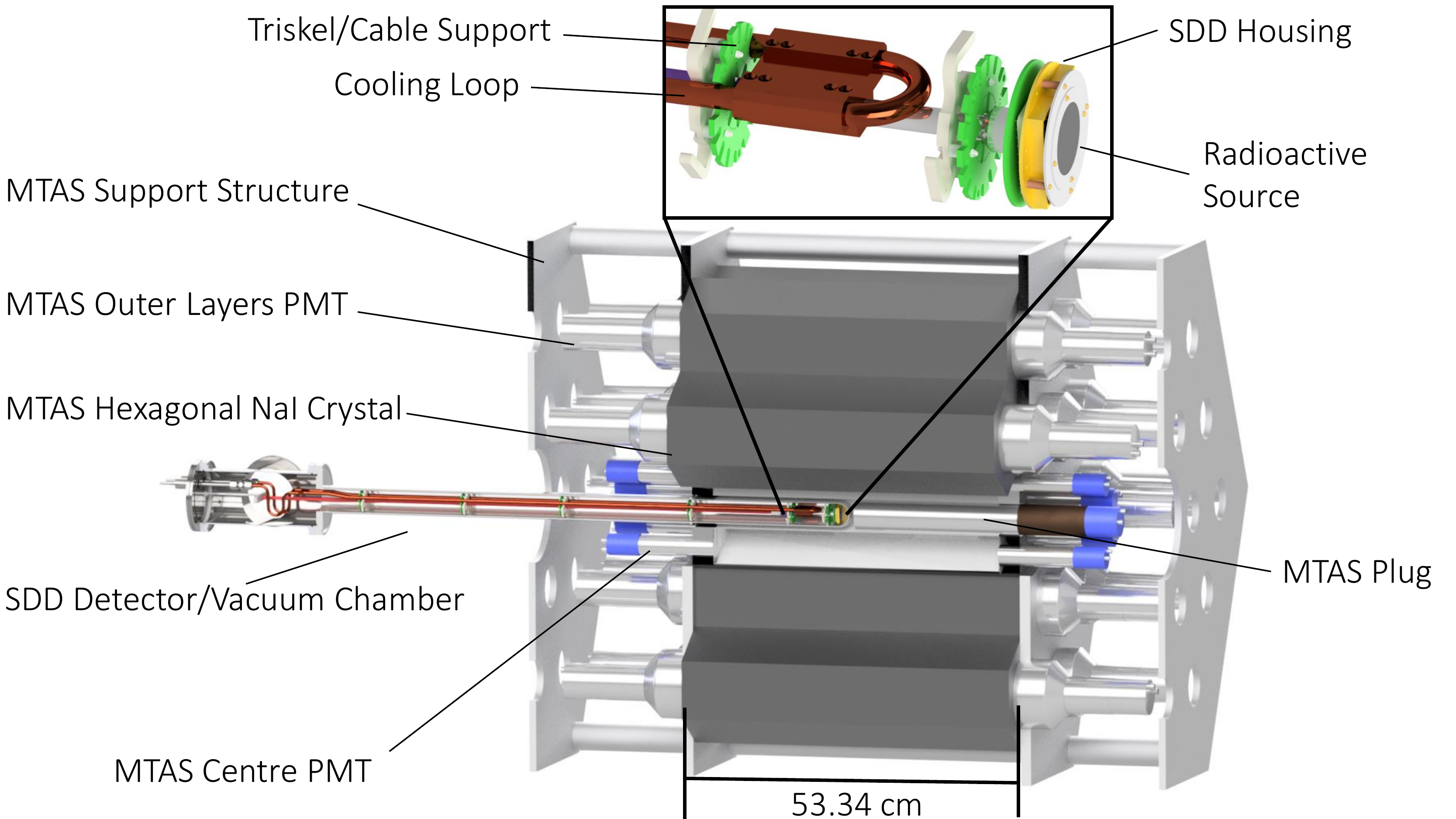}
	\caption{\label{Fig:KDK_Experiment_Total.png}Full KDK SDD/MTAS Experiment. Close up on SDD housing, radioactive source holder and cooling loop. }
\end{figure*}

The complete cooling setup is placed inside a custom designed vacuum tube, shown in Fig.~\ref{Fig:KDK_Experiment_Total.png}. The tube, and its internals, serve to centre the detector and the source in MTAS reproducibly. The two flanges not shown are an electronic feedthrough flange, containing a Sub-D 25 pin connector and three BNC feedthrough connections, and an ISO-100 to KF-40 adapter that connects to a turbo pump setup. 
The vacuum tube has a 30~cm-long aluminum cap with a hemispherical end. The inner diameter of the cap is 4.85~cm with a wall thickness of 0.63~mm. This cap was constructed by Wejay Machine Productions Co. Ltd, Canada with welding performed by Laser Weld Creation, Canada. 
 The cap is attached to a larger (68.8~cm) aluminum tube with the same inner diameter but a wall thickness of 2.31~mm. The resulting ensemble (known as the MTAS insert) is wrapped with a plastic collar that ensures a source resting in front of the SDD will consistently reside in the centre of MTAS. Small, cylindrical copper spacers make certain that the source is 1~mm from the surface of the SDD and an aluminum rod was used to align the SDD/source to the midpoint of the hemisphere of the insert. Support triskels are also placed evenly inside the vacuum chamber to manage the cables from the SDD and centre it radially.

\subsubsection{\label{subsubsec:K40Source}\texorpdfstring{\K\ source development}{K-40 source development} }

Specifications for the \K\ source for the composite method are that it should contain $\sim 10^{18}$ atoms of \K\ to ensure sufficient signal, and that it be thin enough ($\lesssim 10 \ \mu$m) to limit self-absorption of the 3~keV X-rays.
This is implemented as a disk of 1~cm diameter, similar to that of the SDD.  The source rests on a round graphite substrate of 23~mm in diameter and 0.23~mm thick. 

Two approaches were explored for the creation of the source: ion implantation and thermal deposition. The implantation of \K~ions into the graphite substrate used the on-line test facility (OLTF) at ORNL~\cite{beene_isol_2011}. A surface ionization ion source was used to produce singly charged K ions from enriched KCl with a 3.15$\%$ \K~content. The ions were extracted from the ion source at 20~kV, transported through a 90-degree dipole magnet optimized for A=40 mass selection, and then focused toward the graphite substrate with an implantation energy of 20~keV. The substrate was placed inside a shielded Faraday Cup (FC) by which the total incident \K~ion current was measured.  The FC consisted of an entrance aperture of 1~cm diameter to limit the implanting \K~beam size to be about the same as the source geometry. The predicted advantage of this method was that the only radioactive source would be \K~atoms. However, it was found that only $\sim 10 \%$ of the total implanted activity was in the graphite substrate, the rest was on the interior of the FC. This suggested that self-sputtering would prevent the source from reaching the required activity. Additionally, this method produced a noticeable $^{125}$Sb contamination in the source, the cause of which was unknown. For the above reasons, ion implantation was not used for the creation of the experimental source.

\begin{figure}[ht]
  \centering
  \includegraphics[width=1.0 \linewidth]{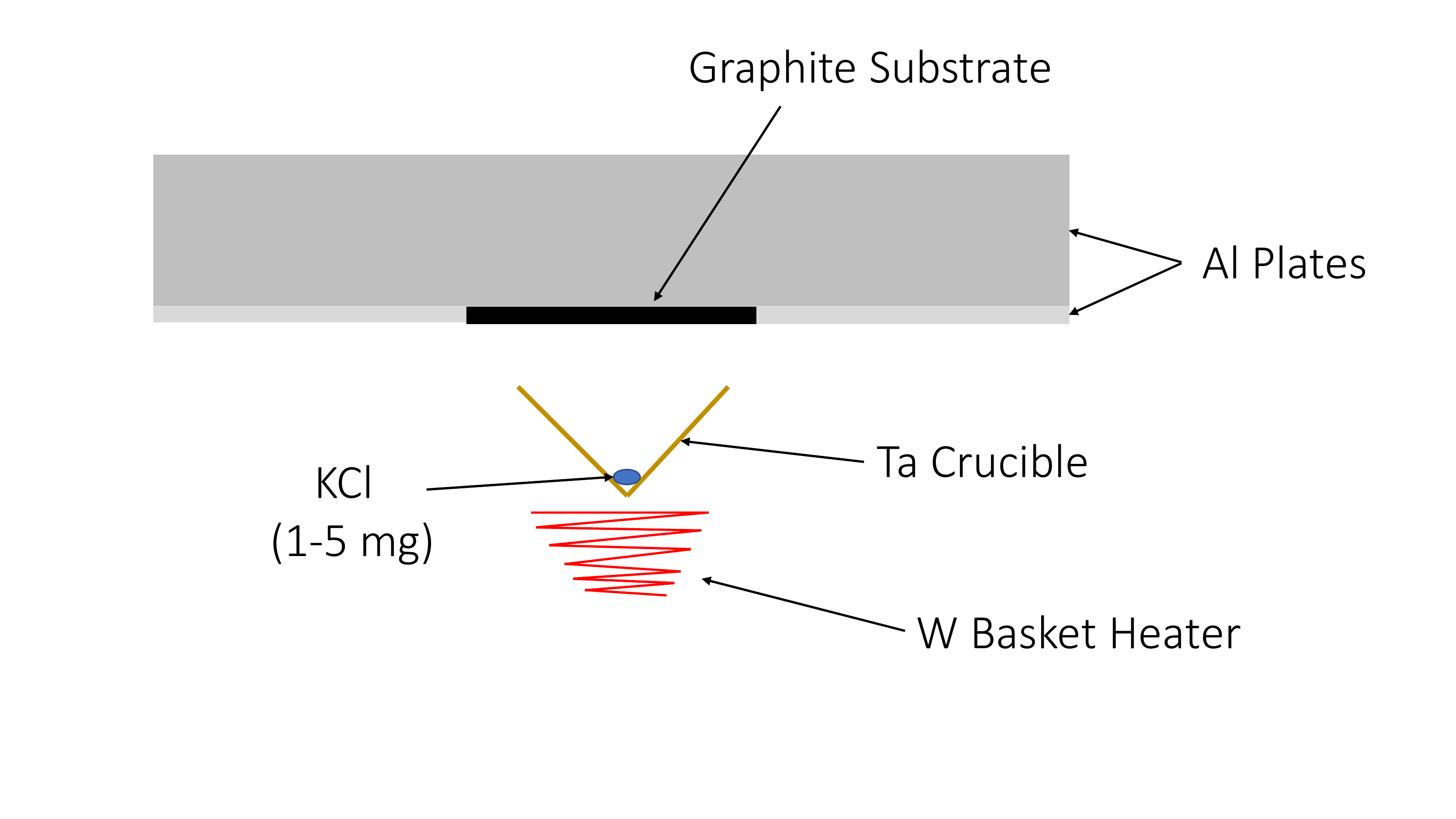}
	\caption{\label{Fig:KDK_Source_Creation.pdf} Schematic drawing of the KCl deposition setup. For clarity, the tantalum crucible is pictured outside of the tungsten basket heater in which it  resides.}
\end{figure}

Instead, thermal deposition was used to place a thin KCl film on the graphite substrate. Assuming KCl molecules are uniformly distributed in the source disk of 1~cm diameter, the desired activity can be attained with a film thickness of about about 25, 8, and 5~$\mu$m for 3.15$\%$, 10$\%$, and 16$\%$ enrichment of \K, respectively. Thermal deposition was carried out in a high vacuum evaporation system with a specially designed thermal evaporator at the Center for Nanophase Materials Sciences (CNMS) at ORNL.  The special evaporation/deposition assembly was constructed with the configuration as illustrated in Fig.~\ref{Fig:KDK_Source_Creation.pdf}. The conical-shaped crucible, made from a thin Ta metal sheet with base diameter of about 12~mm and cone height of 4~mm, centred the KCl feed material. The crucible was heated with a tungsten wire basket heater of 9~mm inner diameter and 13~mm height from Ted Pela, Inc. USA. The basket heater was resistively heated by an electrical current, which directly controlled the temperature. The graphite substrate was sandwiched between two aluminum plates and held at approximately 2 to 3~mm above the crucible. The confined space between the crucible and the substrate allowed a substantial fraction of the evaporated material to hit the substrate, but due to this close geometry the substrate was also heated to high temperatures, decreasing deposition efficiency. 
%Therefore, a rapid evaporation process (higher crucible temperature for a shorter time) was used to prevent the undesired overheating of the substrate.

%
%

\begin{figure}[ht]
  \centering
  \includegraphics[width=1.0 \linewidth]{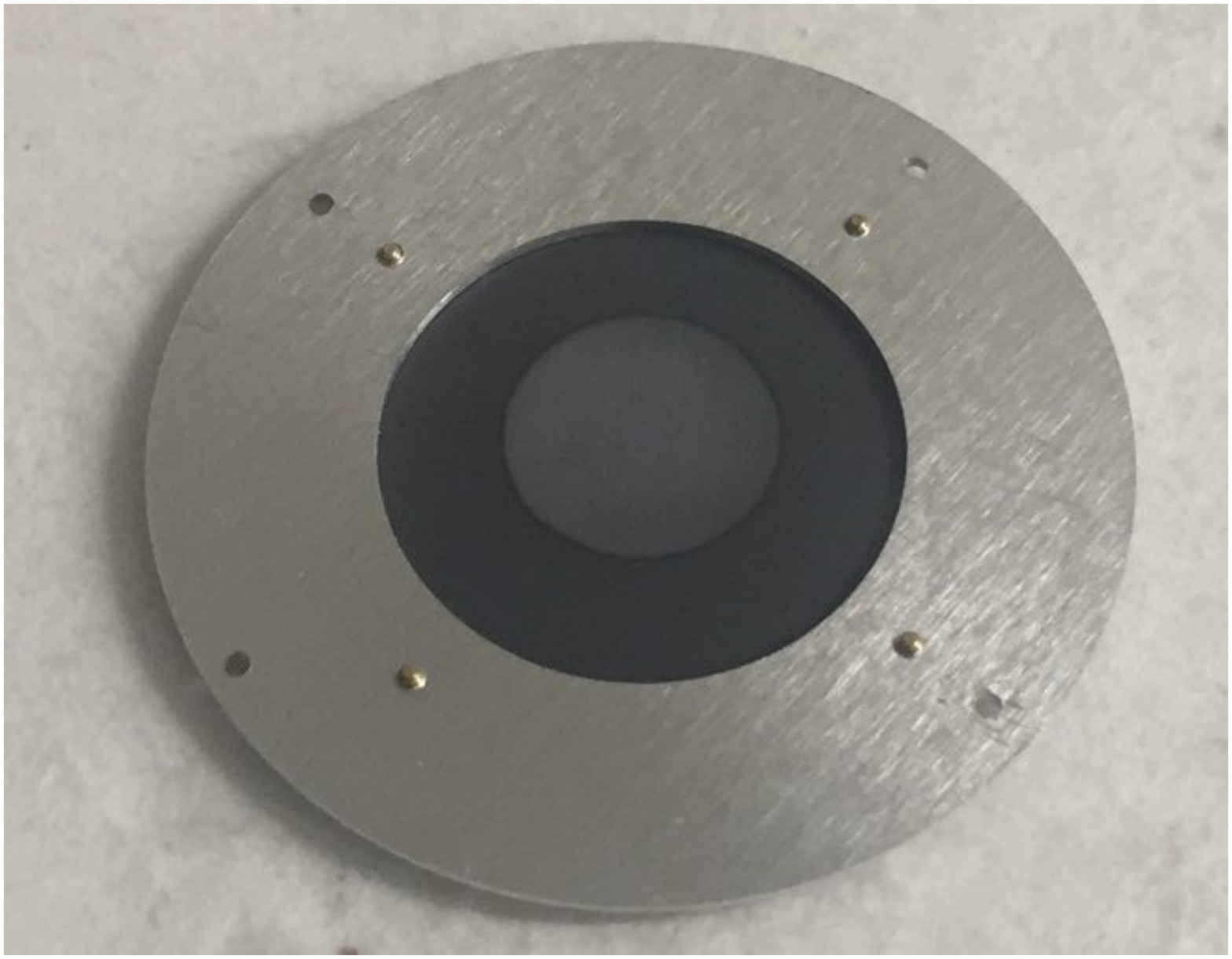}
	\caption{\label{Fig:KDK_Source_Image.JPG} The \K~thin film source on graphite substrate in the KDK detector holder, made from 16.1(6)$\%$ enriched KCl with $\sim 9 \times 10^{17}$ \K~atoms.}
\end{figure}

To compromise between decreased efficiency due to slow heating and the risk of sputtering due to fast heating, tests with natural KCl powder were carried out to find the optimal deposition parameters.
%
%Tests with natural KCl powder were carried out to find the optimal deposition parameters, a compromise between decreased efficiency due to slow heating, and the risk of sputtering due to fast heating. 
%
By varying the deposition time and the heater current, we found the best condition was a heating time of 2~minutes at a heater current of about 18.9~A, which was the maximum heating current of the evaporator, under a vacuum of 10$^{-6}$ to 10$^{-5}$~Torr. Under this condition, feed materials up to 5~mg KCl powder in the Ta crucible could be completely evaporated. After the 2-minute heating, the heater current was immediately turned off to allow the substrate to cool naturally. We were successful in depositing a smooth layer of KCl on graphite disks reproducibly, as shown in Fig.~\ref{Fig:KDK_Source_Image.JPG}.  The substrates were weighed before and after deposition to determine the deposition efficiency (the ratio of deposited KCl to feed KCl) and it was in the range of 30-50$\%$. 

\begin{figure}
\centering
\includegraphics[width=1.0\linewidth]{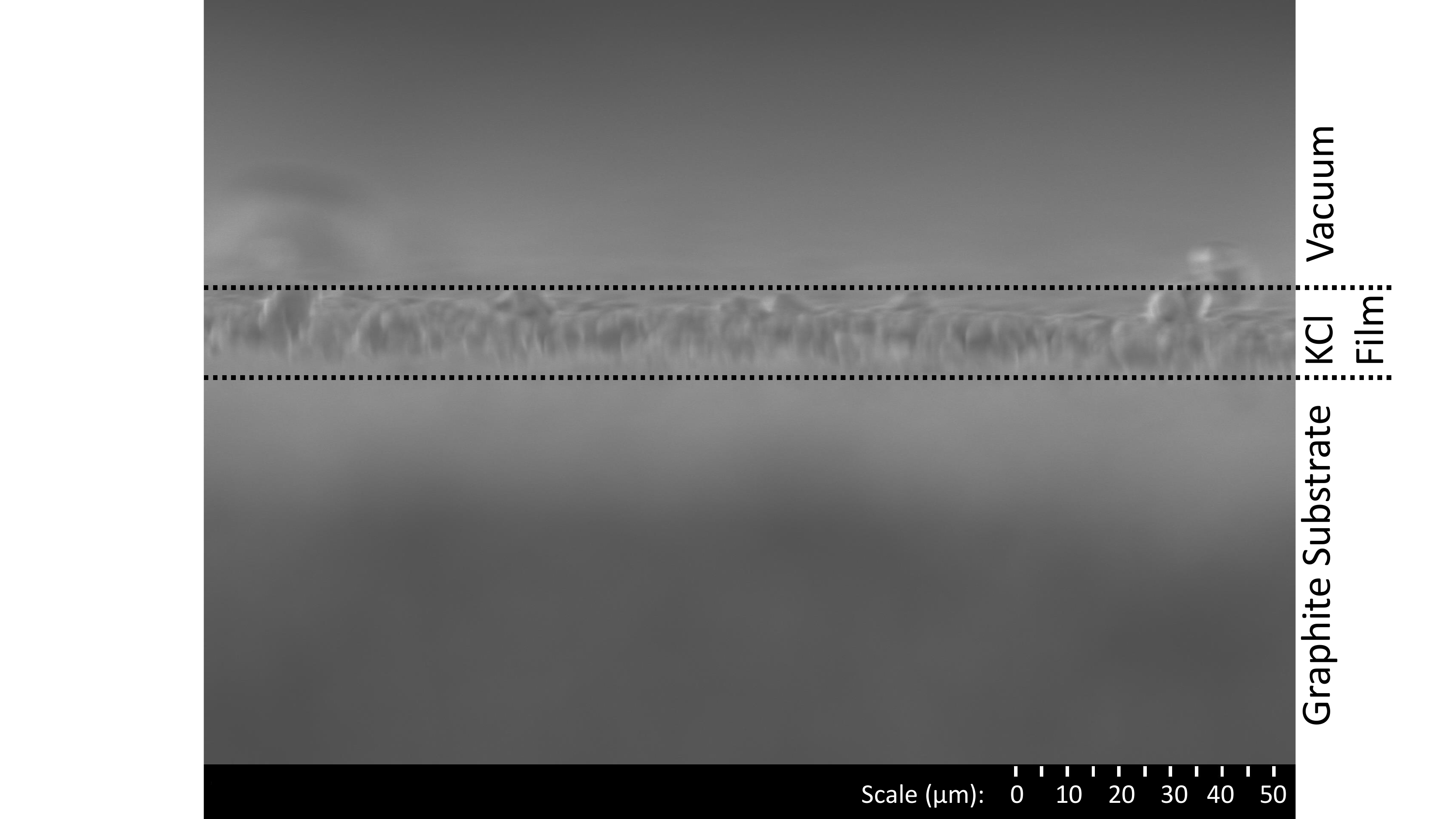}
\caption{\label{Fig:KCl_Source_SEM.pdf} SEM image of the cross-section of the natural KCl film sample for thickness estimation. 10 measurements were made per image. A total of 20 measurements gave an average thickness of 11(1)~$\mu$m.  }

\end{figure}

The thickness of the KCl films produced was characterized by scanning electron microscopy (SEM) analysis. Two KCl films were analyzed, one with 1.5(1)~mg of natural KCl and the other with 1.45(12)~mg of 3.15$\%$ enriched KCl. The thickness of the natural KCl film was measured to be 11(1)~$\mu$m (Fig.~\ref{Fig:KCl_Source_SEM.pdf}). It is noted that the calculated thickness of 1.5~mg KCl uniformly distributed in a circle of 1-cm diameter is 9.7(6)~$\mu$m, which is statistically consistent with the SEM measurement. The ratio of the calculated thickness to the SEM measurement is 0.9(1).

The final KCl thin film was made with 16.1(6)$\%$ enriched \K. The deposition used 1.76(7)~mg of the enriched KCl as feed material and a net of 0.69(11)~mg KCl was deposited on the graphite substrate, corresponding to a deposition efficiency of about 39$\%$. Based on how the source was prepared, the total activity in the source should be about $9\times 10^{17}$ atoms of \K. 
%
%The calculated film thickness is 4.44(71)~$\mu$m. 
%The \K~source thickness is estimated to be 5.1(9)~$\mu$m.
%
Using the mass of deposited enriched KCl, and scaling by the ratio of SEM thickness to calculated thickness for the natural KCl film, the final \K~source thickness is estimated to be 5.1(9)~$\mu$m.

Determining the actual activity of the source by measuring decays is complicated.  For instance, the gammas from \ECStar\ decays are negligible compared to the background in MTAS, and measuring the lower energy quanta like betas or X-rays depends on knowing the efficiency of the SDD and the self-absorption of the source.

\subsection{\label{sec:Detec_Charac_KSI}Homogeneous inner detector: KSI scintillator}

We have also explored the use of the scintillator \KSI~(KSI)~\cite{stand_potassium_2013} as a combined \K\ source and X-ray detector. The light yield of KSI varies depending on the packaging and crystal size but has been reported to be as high as 94,000 photons/MeV at 662~keV with good proportionality compared to other scintillators~\cite{stand_stracuzzi_discovery_2018}. Grown with natural potassium and with an estimated density of 4.39~g/cm${}^3$, the intrinsic activity of KSI is $\sim$6.5~Bq/cm${}^3$.

\begin{figure}[hb]
  \centering
  \includegraphics[width=1.0\linewidth]{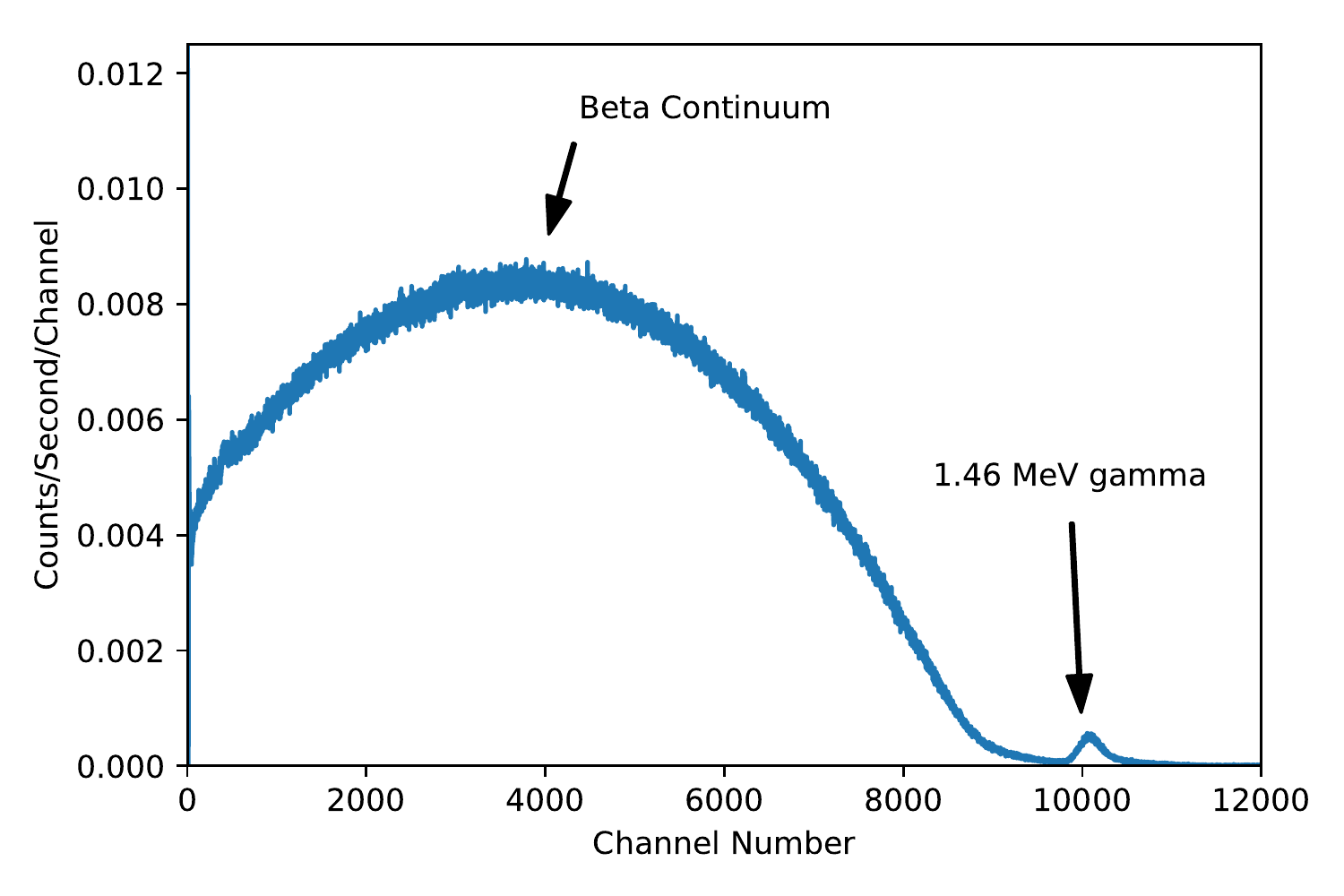}
	\caption{\label{Fig:KSI_Internal_rate.pdf}Measurement of a shielded 2.54~cm diameter by 2.54~cm long KSI scintillator.  Main features are the 1.46~MeV gamma line as well as the continuous beta spectrum, both from \K.}
\end{figure}

In Fig.~\ref{Fig:KSI_Internal_rate.pdf}, the   spectrum of an approximately 2.54~cm diameter by 2.54~cm long cylindrical KSI scintillator is provided. 
The measurement was carried out in a lead cabinet to shield the scintillator from external radiation, and is dominated by the internal \K.
The 1.31~MeV end point beta continuum and 1.46~MeV $\gamma$-ray emitted during the decay of \K\ are clearly observable in the pulse height spectrum. Note that Sr would contribute to the beta continuum but with a much lower activity.  

\begin{figure}[ht]
  \centering
  \includegraphics[width=1.0\linewidth]{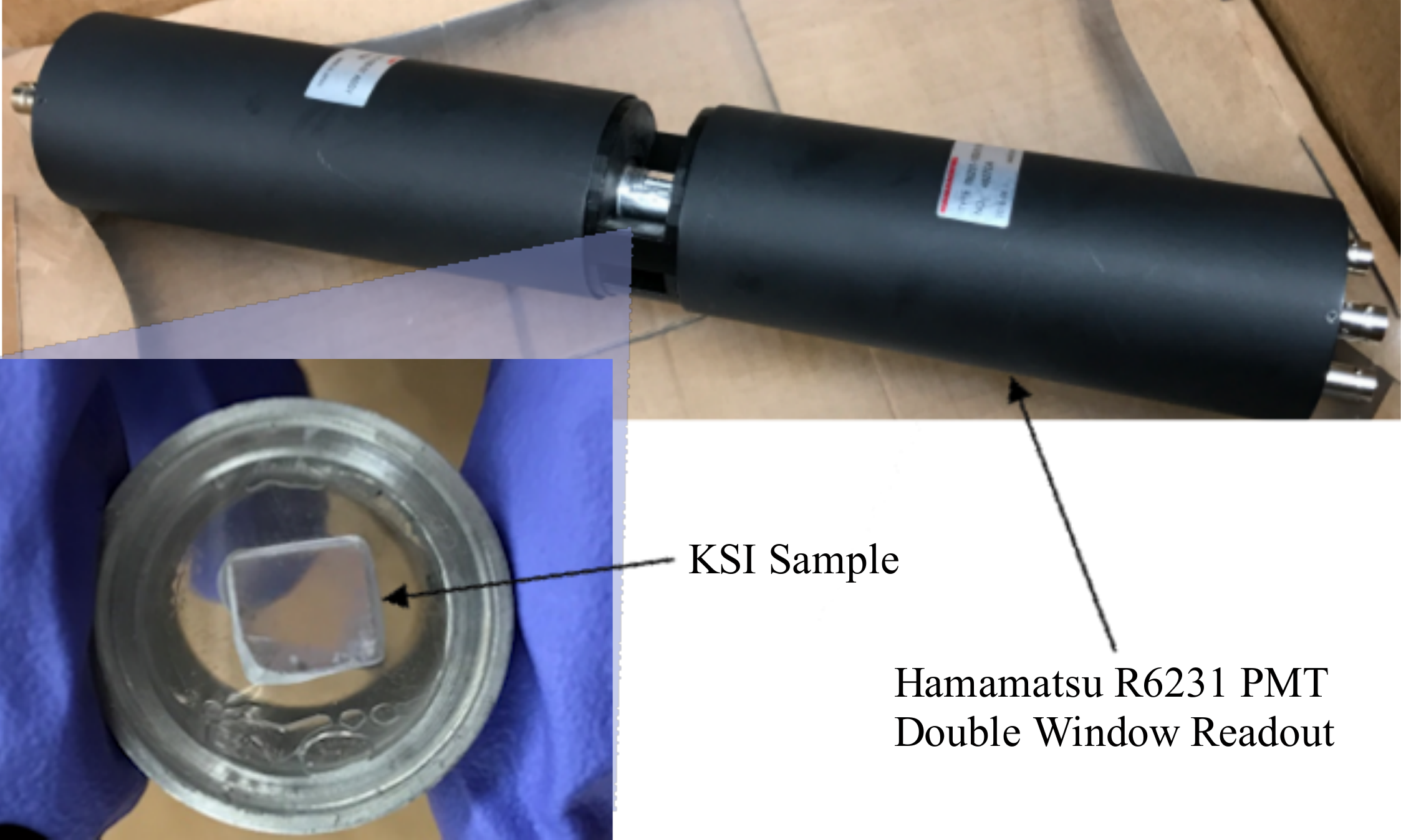}
	\caption{\label{Fig:KSI_combined_pmt_aluminum.pdf}$7\times7\times19.9 \mbox{ mm}^3$  rectangular KSI sample wrapped in 400 $\mu$m of teflon sealed inside an aluminum housing with a nitrogen atmosphere placed in the centre with a custom 3D printed polyethylene bracket holding the setup together.}
\end{figure}

\begin{figure}[ht]
  \centering
  \includegraphics[width=1.0\linewidth]{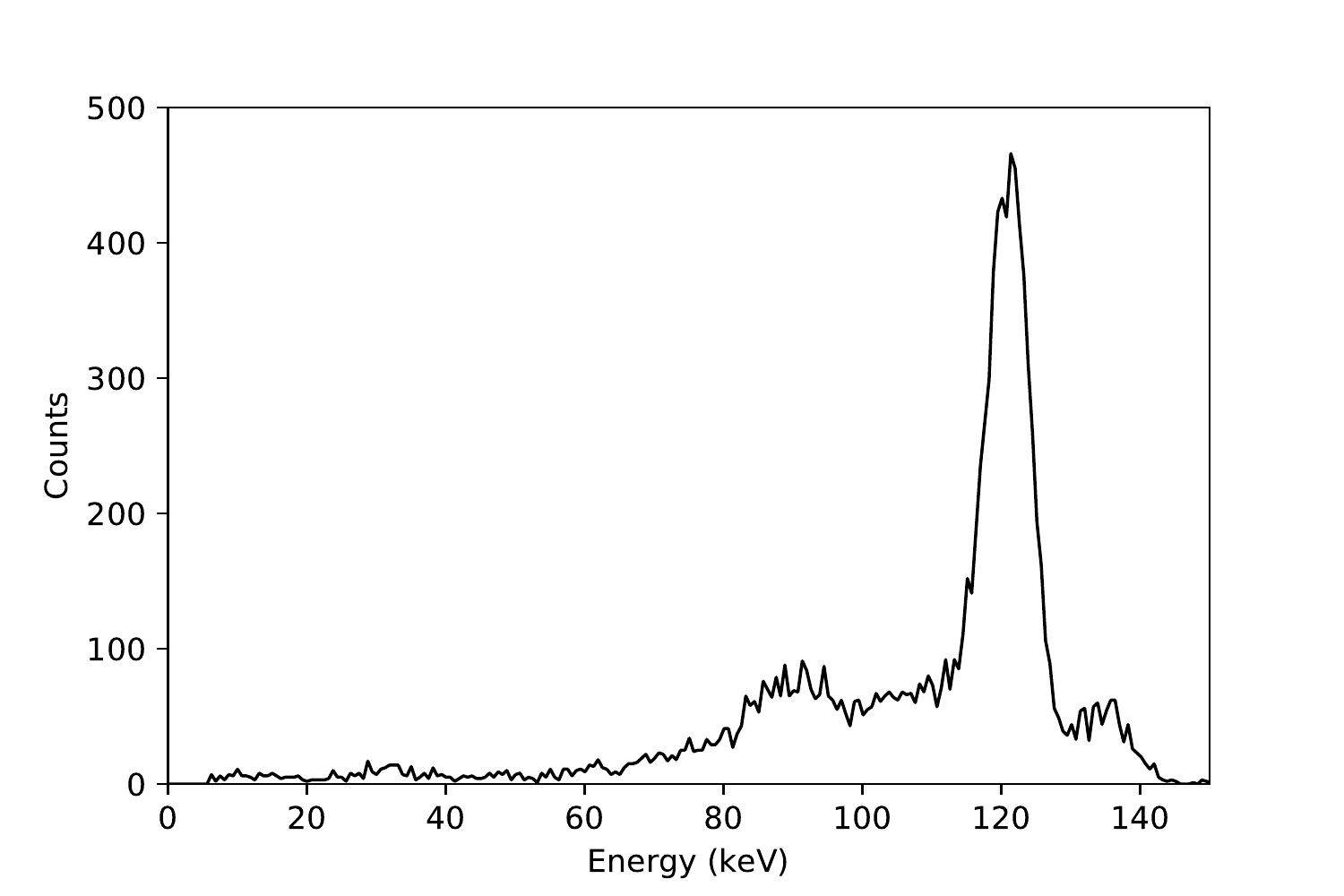}
	\caption{\label{Fig:KSI_gain_match.pdf}Analog sum of both PMTs when the KSI scintillator is exposed to a ${}^{57}$Co source.}
\end{figure}

The KSI sample used for the KDK measurements was grown using the vertical Bridgman technique as described in~\cite{stand_potassium_2013}. The crystal was cut and polished into a $7\times7\times19.9\mbox{ mm}^3$ rectangular parallelepiped. The sample was then wrapped in an approximately 400~$\mu$m-thick layer of teflon inside of a 2.5~mm thick aluminum housing with a nitrogen atmosphere. The aluminum housing contained a small, 2~mm diameter hole covered with a 50~$\mu$m aluminized mylar film to allow in low energy ionizing radiation for calibration purposes  with windows on both ends for a double readout, shown in Fig.~\ref{Fig:KSI_combined_pmt_aluminum.pdf}. The crystal was mounted to each silicon oxide window using the Eljen optical cement EJ-500. Two R6231 Hamamatsu PMTs were used to readout each end and were held together with a custom, 3D printed bracket. The combined system, shown in Fig.~\ref{Fig:KSI_combined_pmt_aluminum.pdf}, was placed inside the middle of MTAS and utilized the same readout system. In order to process the information from the double window readout, the signals from each PMT must be summed to collect all the light generated during an ionizing radiation event within KSI, shown in Fig.~\ref{Fig:KSI_gain_match.pdf}.

\begin{figure}[ht]
  \centering
  \includegraphics[width=1.0\linewidth]{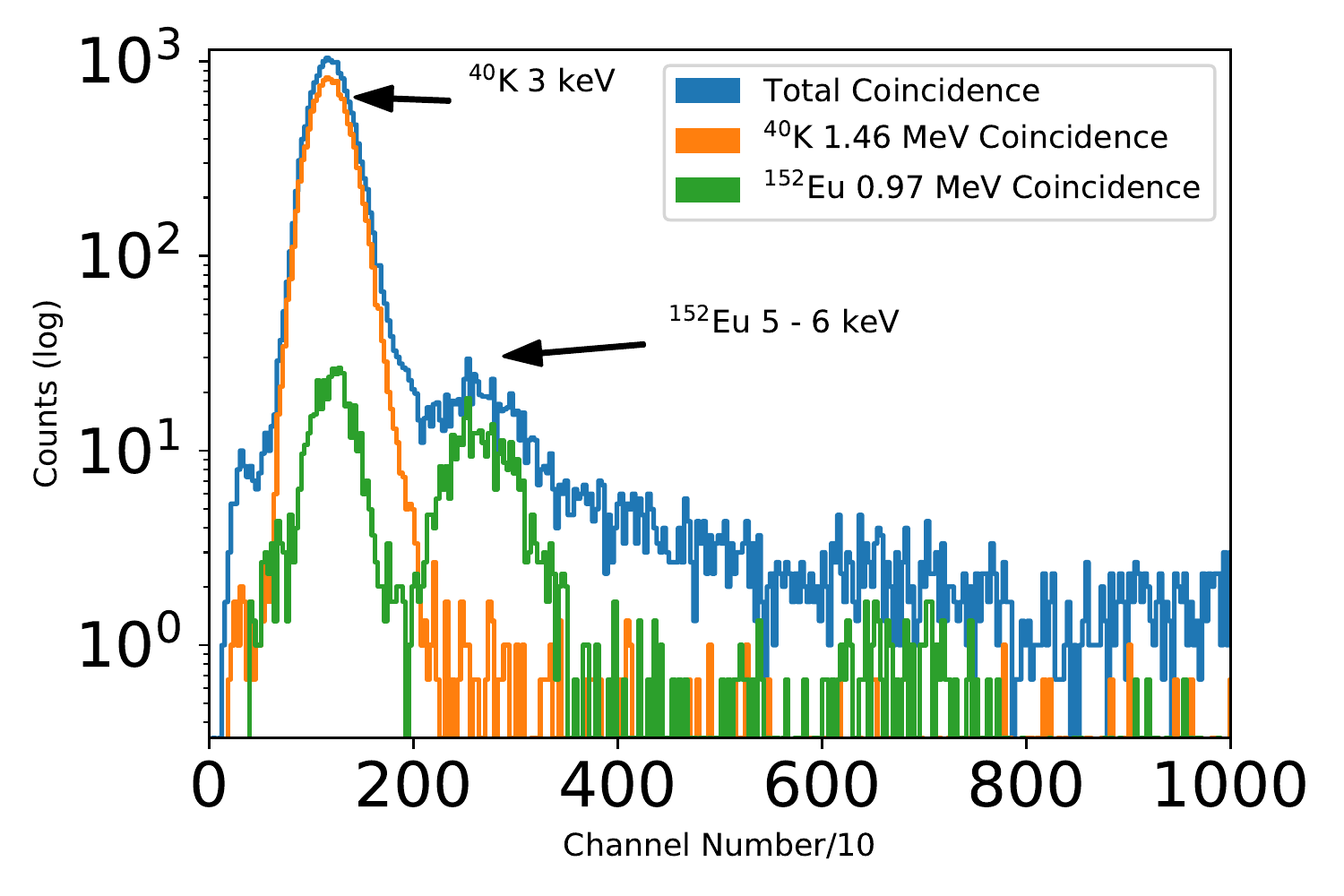}
	\caption{\label{Fig:Eu_activation_Plot.pdf}Spectra from an activated KSI:Eu sample, in coincidence with MTAS. Blue: strictly positive energy in MTAS, showing both $\approx 3$~keV \K\ X/Augers, and Eu X/Augers. Orange: energy in MTAS $\approx 1.46$~MeV, showing mainly X-Auger from \K. Green: energy in MTAS $\approx 0.97$~MeV, showing mainly Eu  where \K\ is less visible. }
\end{figure}

To confirm that the 3~keV X-ray/Auger peak for \K\ is visible in KSI, characteristic features from activation of Eu were used to verify the peak location in this low energy range. The KSI sample was exposed to a 26.7~milliCurie PuBe source for 9~hours and then placed into MTAS so a spectrum could be gathered from the activated Eu content while also tagging the coincident higher energy gammas from the Eu decay. $^{151}$Eu ($n$,$\gamma$) $^{152}$Eu produces photons and Auger electrons with a wide range of energies in the region we are investigating.  $^{153}$Eu ($n$,$\gamma$) $^{154}$Eu also produces the same spectral features, however the nuclear cross section is orders of magnitude lower. \I\ is also activated in this process, but with a 24.99(2) minute half life, the contribution is negligible. This activation method focused only on events from the $^{152}$Eu that were coincident in KSI and MTAS. By gating the positive events in MTAS on the 1.46~MeV gamma from \K\, the contribution of the 5--6~keV X-ray/Auger electrons from Eu are removed leaving only the 3~keV peak, as seen in Fig.~\ref{Fig:Eu_activation_Plot.pdf}. For comparison, the same is done on a characteristic $\sim$970~keV gamma from Eu, reducing the intensity of the 3~keV \K\ peak to those coincident with part of the Compton continuum from the 1.46~MeV. This corroborates that the  peak seen in this low energy spectrum is in fact the 3~keV X-ray/Auger peak. 

Currently, for the purpose of the KDK \K\ measurement, use of the KSI scintillator is limited by our simulation model of the PMT structure, which prevents us from determining the precise tagging efficiency for $\gamma$-rays, as the simulations are sensitive to the distribution of material near the KSI source. Variations in R6231 PMT materials that fall within design and manufacturing parameters result in insufficient precision in the model. An option under consideration is to replace the PMTs with smaller SiPMs. These smaller SiPMs can be sized to cover only the window of the KSI sample, see Fig.~\ref{Fig:KSI_combined_pmt_aluminum.pdf}, leading to a large reduction in mass immediately near the windowed sample. Additionally the design of SiPMs being mostly silicon, glass, and a fiberglass PCB, along with their small size, greatly reduces the amount of material and variation within the structure to be placed into MTAS.

\section{\label{sec:Detec_Charac}Characterizing the SDD in MTAS}

Of the two methods developed, the composite one is currently better understood, and is the focus of the rest of this work.

\subsection{\label{subsubsec:SDD_Data_Reduction}Data reduction and SDD operation}

\begin{figure}[ht]
  \centering
  \includegraphics[width=1.0\linewidth]{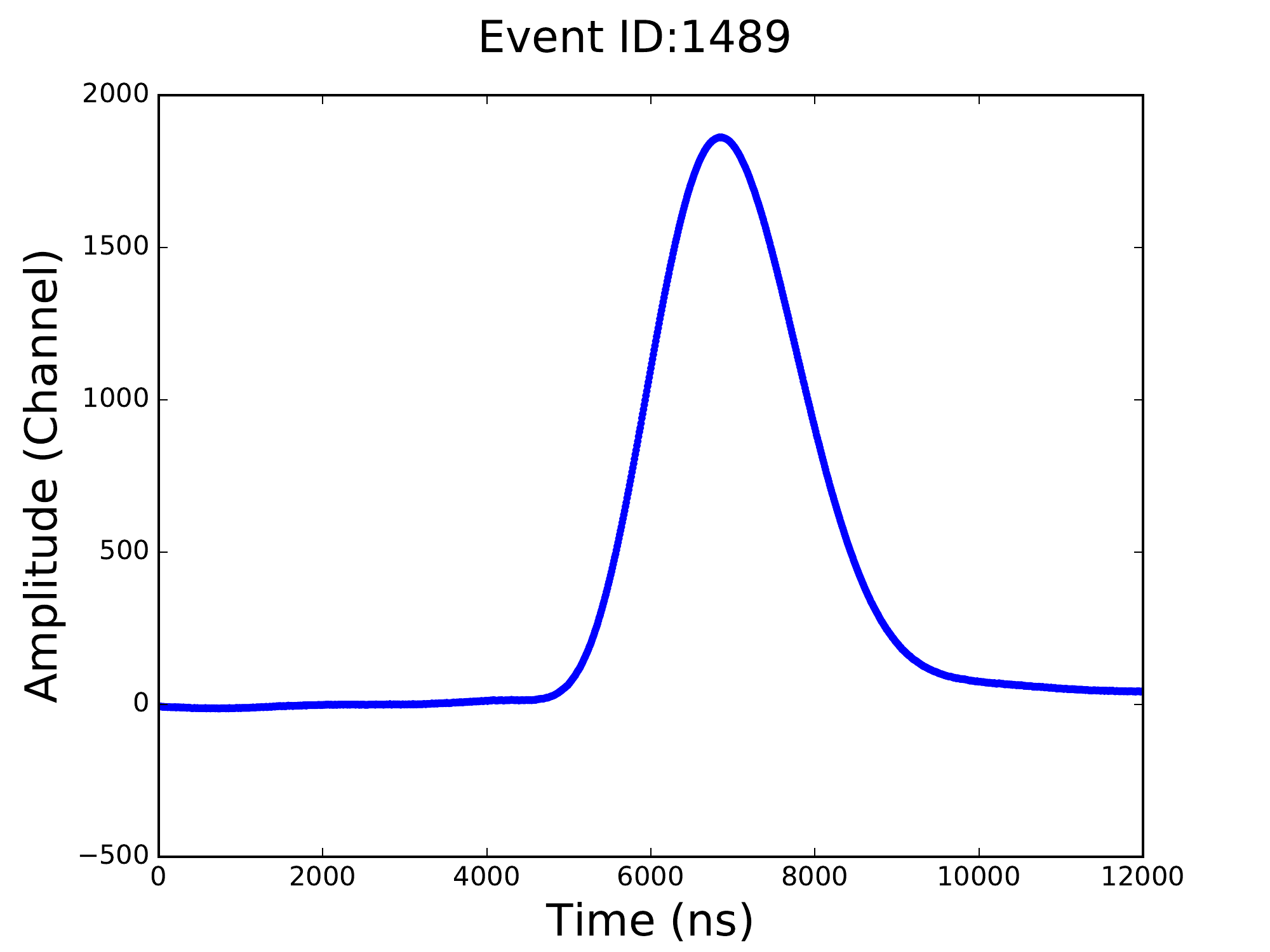}
	\caption{\label{Fig:Pulse_1489_Paper.eps}Shaped SDD pulse for a 8.04~keV X-ray from a \Zn~source. The trace is baseline subtracted. There is a 40$\%$ pre-trigger length. }
\end{figure}

When an event triggers the SDD it is digitized by the same DAQ as discussed in Section~\ref{subsec:MTAS}.  The time window of an SDD trace is 12~$\mu$s with 10~ns sampling and a 40$\%$ pre-trigger length. Based on event rates for most sources in the SDD (Table~\ref{tab:Rate_Table}), the expected number of spurious coincidences in an SDD window is $\lesssim 5\times 10^{-4}$. Baseline subtraction of the event is done during the offline analysis by averaging the y-axis data of the pre-trigger and subtracting the value from the trace. The amplitude of the event is determined by fitting a gaussian to the trace. The amplitude corresponds to the energy of the event and the energy calibration is discussed in Section~\ref{subsec:Calibration}. The standard deviation of the baseline is used as the uncertainty on each sample of the trace for the fit. A sample trace for an 8.04~keV \Zn~event can be seen in Fig.~\ref{Fig:Pulse_1489_Paper.eps}.

\begin{figure}[ht]
  \centering
  \includegraphics[width=1.0\linewidth]{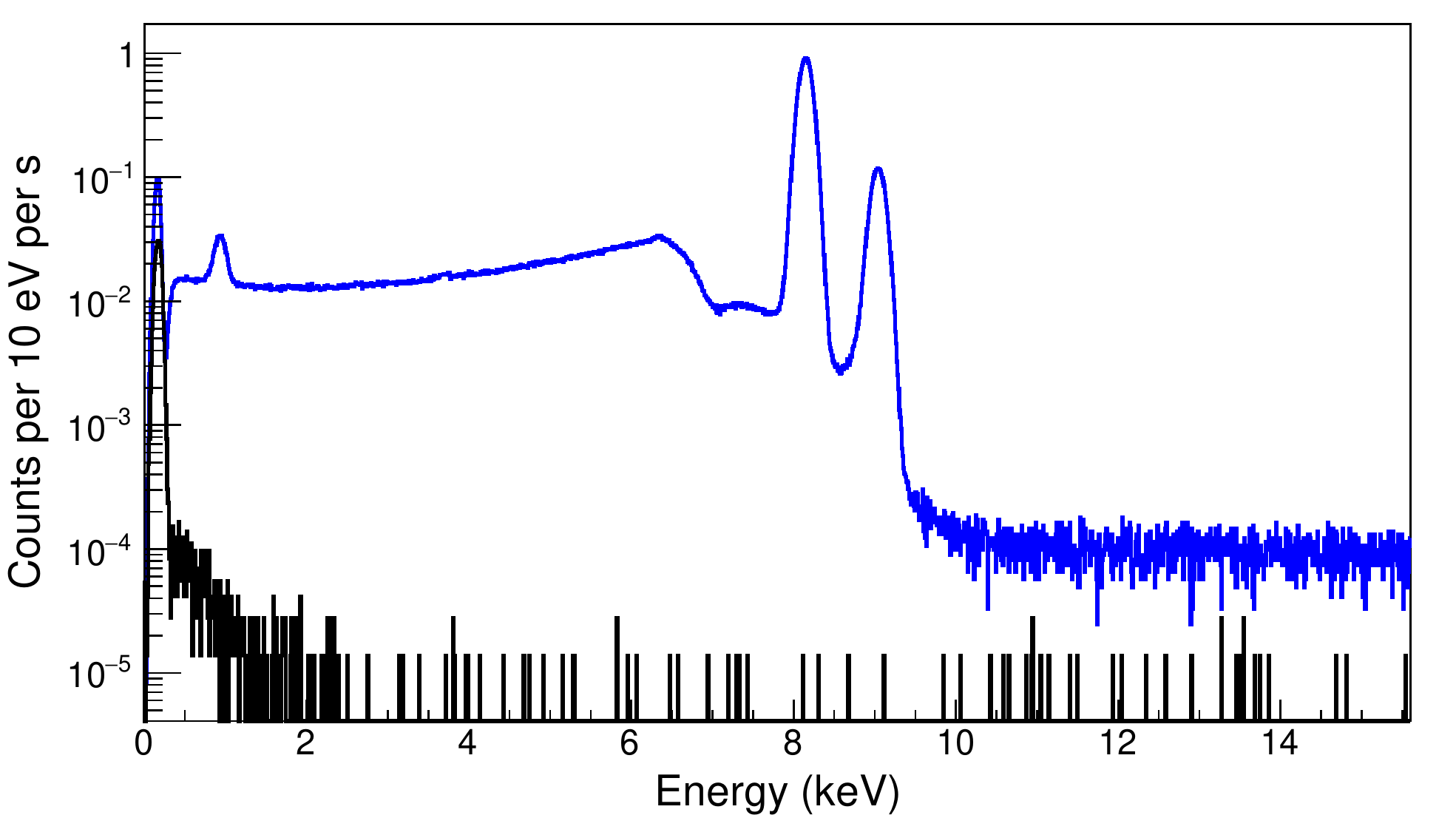}
	\caption{\label{Fig:Zn65_vs_BKG_Spectrum} SDD spectra with no condition on MTAS.  Blue: \Zn. Black: background (i.e. no source present). Energy resolution is 198~eV~FWHM at the \Zn \ \Kalpha\ X-ray peak, and the 5$\sigma$ noise threshold is 370~eV. A description of the different components of the \Zn\ spectrum is offered in the text. }
\end{figure}

All traces produced from the SDD are fit in this manner and each run performed was monitored for gain stability. The information is then combined with the MTAS energy and timing data for further offline analysis. A global rolling window (as described in Section~\ref{subsec:MTAS}) is used to determine coincidence between the SDD and MTAS. A sample spectrum for the SDD using a \Zn\ source is shown in Fig.~\ref{Fig:Zn65_vs_BKG_Spectrum}, a resolution of 198~eV FWHM at 8.04 keV, and a 5$\sigma$ noise threshold of 370~eV was achieved. The decay scheme of \Zn\ can be found in~\cite{be_table_2006}. The expected X-ray energy values quoted, can be found in Table~\ref{tab:SDD_Energy_Calibration}. 

In the spectrum the \Zn~L:~0.93~keV, \Kalpha:~8.04~keV and \Kbeta:~8.94~keV X-ray peaks are visible. Also visible are the \Zn~Auger electrons which are identifiable by their characteristic wide continuous energy spectrum below the X-ray peaks. Unlike X-rays, the exact energy deposit of Auger electrons depends sensitively on the path out of the source and through the SDD dead layer, thereby forming a continuous energy spectrum instead of a single peak. The background for energies greater than 10~keV is formed by low-energy interactions in the SDD from other \Zn~decay components.

\subsection{\label{subsec:Calibration}SDD energy calibration}

The energy scale of the SDD was calibrated using four sources: \Mn, \Zn,  \Y,  and \K\  (the \K\ signal region remained blinded). The \Mn, \Zn~and \Y~source were purchased from Spectrum Techniques, they are open sources with $<$1~cm diameter. The corresponding run time, energies and shell capture for each source can be found in Table~\ref{tab:SDD_Energy_Calibration}.  The SDD energy values are the weighted average of the quantities found in the most recent Table of Radionuclides publications in~\cite{be_table_2010,be_table_2006,be_table_2004,be_table_2016}. 

\begin{figure}[ht]
  \centering
  \includegraphics[width=1.0\linewidth]{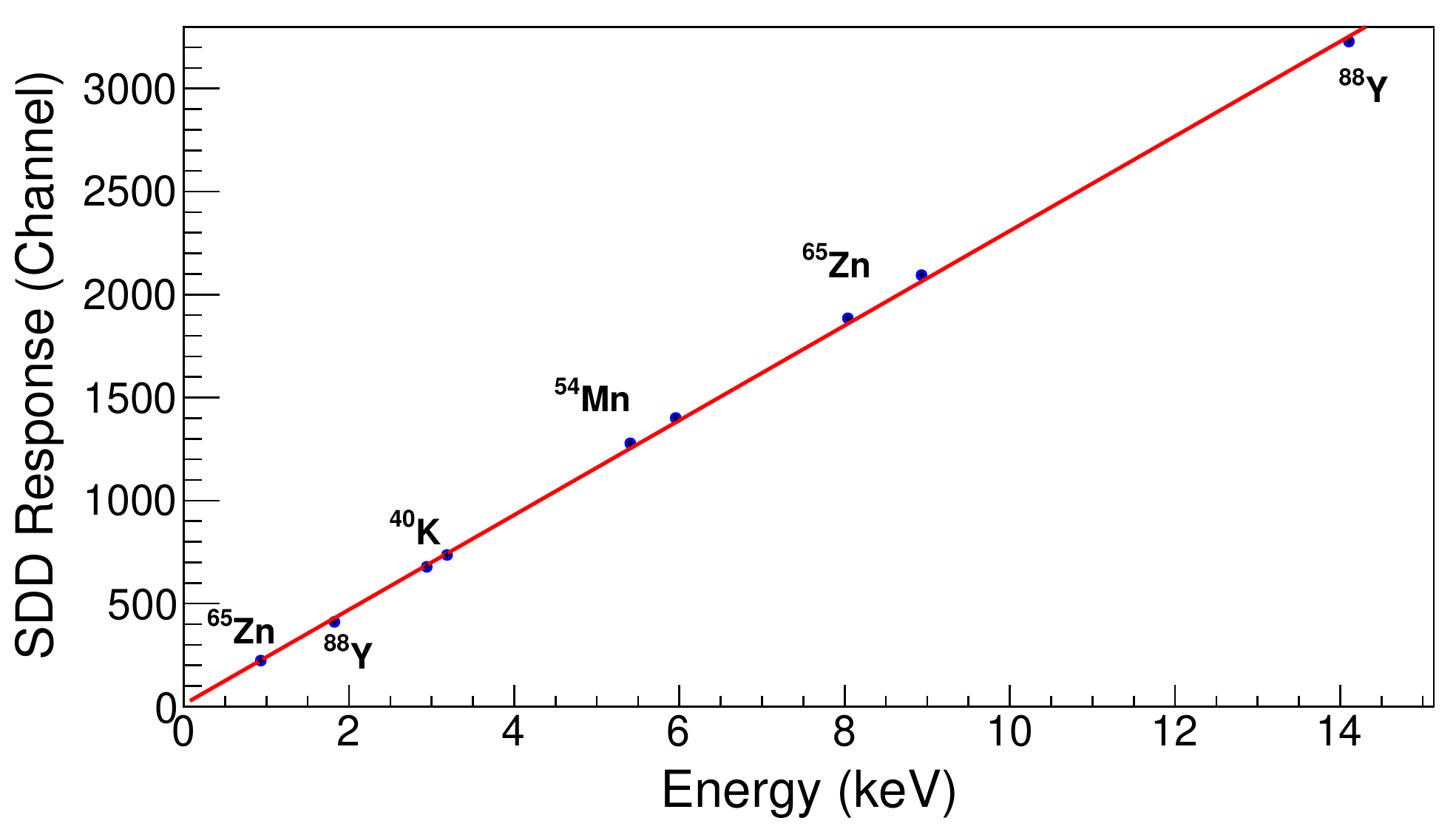}
	\caption{\label{Fig:SDD_Calibration.png}Calibration of the SDD with \Mn, \Zn, \Y~and \K~sources. The slope is 229.8(1)~Channel/keV and the y-intercept is 10.4(2)~Channel.}
\end{figure}

The energy linearity of the SDD can be seen in Fig.~\ref{Fig:SDD_Calibration.png}. The multiple source calibration shows that the SDD easily achieves its target for measuring the \K~X-rays.

\begin{table*}[ht]
\centering
\begin{tabular}{cclcc} 
    
\hline        
Source & Activity & Energies  & Energies  & Run Length  \\
&(kBq) &(X-Ray, eV)&(Gamma, keV)&(Days)\\\hline
\Mn& 0.133(11) & \Kalpha: 5411.68   & 835  & 4.7 \\
 & (2017/02)&\Kbeta: 5966.89&& \\
&&&&\\
\Zn & 0.999(74) &L : 937.1 & 1115 &1.4  \\
&(2017/02)&\Kalpha: 8041.1 && \\
&&\Kbeta: 8941.25 && \\
&&&&\\
\Y & 18.5 &L : 1827.84 & 3584   &0.5 \\
&(2017/05/26)&\Kalpha:  14121.01 & 3218 &\\
&&&2734 & \\
&&&1836 & \\
&&&&\\
\K & 0.015  & \Kalpha: 2957.04 & 1460  & 44  \\
& (2017/08) & \Kbeta: 3190.5 && \\
&&&& \\
No Source && N/A & N/A & 1.3\\\hline

\end{tabular}
\caption{Overview of the KDK experimental data taking campaign.}
\label{tab:SDD_Energy_Calibration}
\end{table*}

\subsection{\label{subsec:Coincidence_Spectrum}SDD-MTAS coincidence spectrum}

\begin{figure}[ht]
  \centering
  \includegraphics[width=1.0\linewidth]{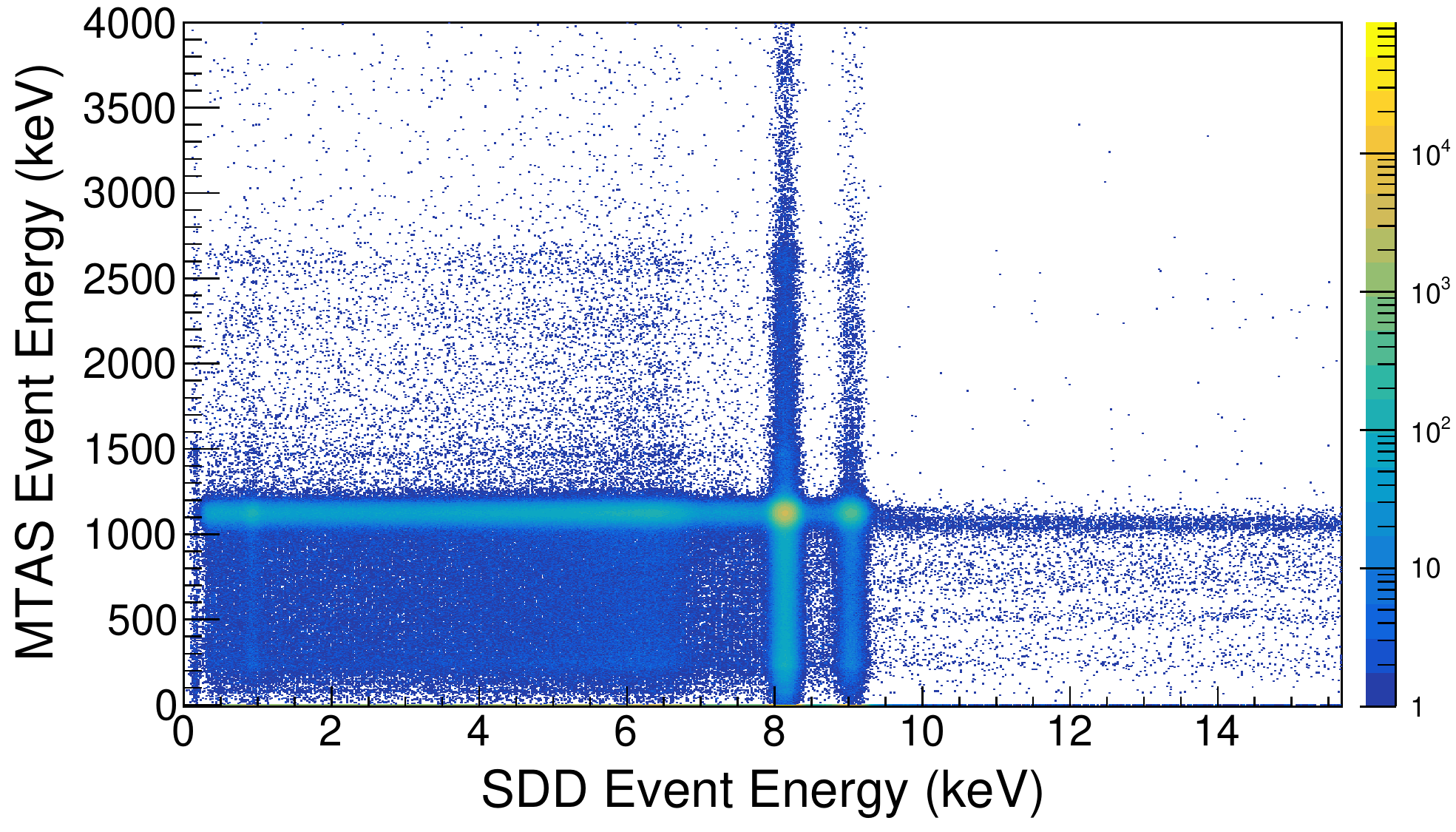}
	\caption{\label{Fig:Zn65_Coincidence_Spectrum.eps}SDD/MTAS \Zn~coincidence spectrum (2~$\mu$s window). Color scale represents counts per bin, with each bin having size 16 keV SDD/1000 $\times$ 4 MeV MTAS/1000 
	(counts per MTAS energy bin per SDD energy bin). The \Kalpha~and \Kbeta~coincident peaks are clearly visible with the description of the horizontal and vertical lines being found in the text. }
\end{figure}

A coincidence plot can be made between the energy recorded in MTAS and the energy recorded in the SDD.  Fig.~\ref{Fig:Zn65_Coincidence_Spectrum.eps} shows the coincidence plot for \Zn. There are two sections to every coincident plot: the coincident  and anti-coincident region. Based on Section~\ref{subsec:MTAS}, if the event in the SDD has no coincident partner it is assigned an MTAS energy  of -1.0~keV and resides in the anti-coincidence region. Any event with a coincident partner is given the corresponding MTAS energy and resides in the coincident region. 

The main feature of these plots is the coincidence peak. This is when the X-ray and $\gamma$-ray are fully captured by their respective detectors. For the \Kalpha\ case of~\Zn\ this corresponds to a 8.04~keV X-ray and 1115~keV $\gamma$-ray. The vertical line below the coincidence peak represents when the $\gamma$-ray is not fully captured (whether  due to Compton scattering or other effects). A horizontal tail is created to the left when an Auger electron interacts in the SDD simultaneously with a 1115~keV $\gamma$-ray in MTAS. A $\gamma$-ray from the source can arrive at the same time as an external background event creating the vertical tail above the coincidence peak, which is further discussed in Section~\ref{subsec:Geant_Simulations} and~\ref{sec:FalseNegative}. Finally, for the \Zn-specific case it is possible for the $\beta^+$ to annihilate inside the SDD, deposit a small amount of energy, and release two 511~keV gammas which are captured by MTAS as an 1022~keV event. This can be seen in the horizontal tail to the right of the coincidence peak.

\section{\label{subsec:Efficiency} Composite method tagging efficiency and sensitivity}
In the isotopes considered in our study, the dominant decay mode from the excited state is via $\gamma$-ray emission, therefore we focus on detecting such emissions to tag \ECStar.

\subsection{\label{subsubsec:Mn_Exp_Eff}\texorpdfstring{\Mn\ experimental efficiency}{Mn Efficiency}}

\begin{figure}[ht]
    \includegraphics[width=1.0\linewidth]{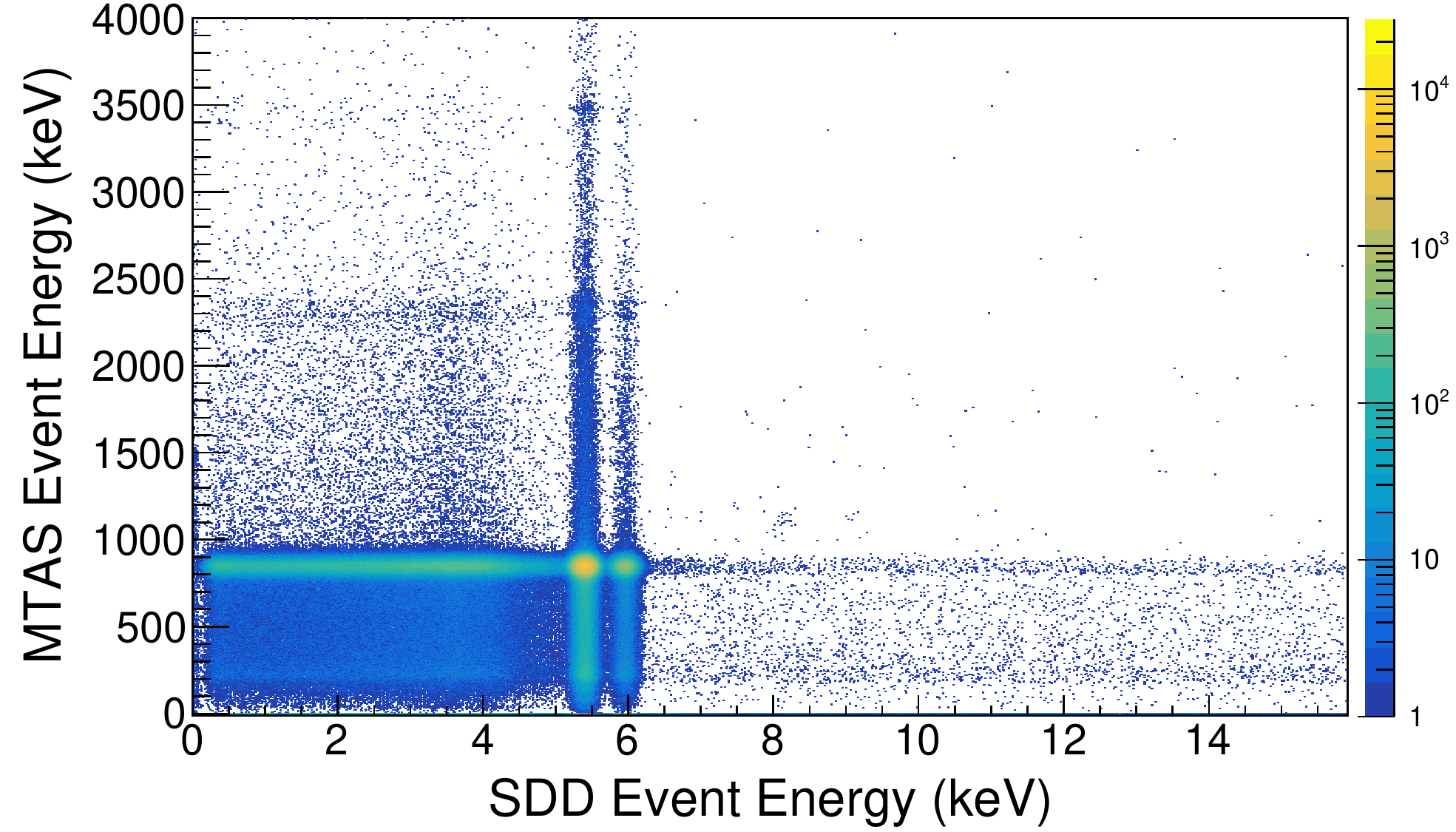}
    \centering
	\caption{\label{Fig:Mn54_Coincident_Plot_2us.eps}SDD/MTAS \Mn\ coincidence spectrum (2~$\mu$s window). Color scale represents counts per bin, with each bin having size 16~keV~SDD/1000 $\times$ 4~MeV~MTAS /1000 
	(counts per MTAS energy bin per SDD energy bin). Clearly visible are the \Kalpha\ and \Kbeta\ coincident peaks. }
\end{figure}

\begin{figure}[ht]
    \includegraphics[width=1.0\linewidth]{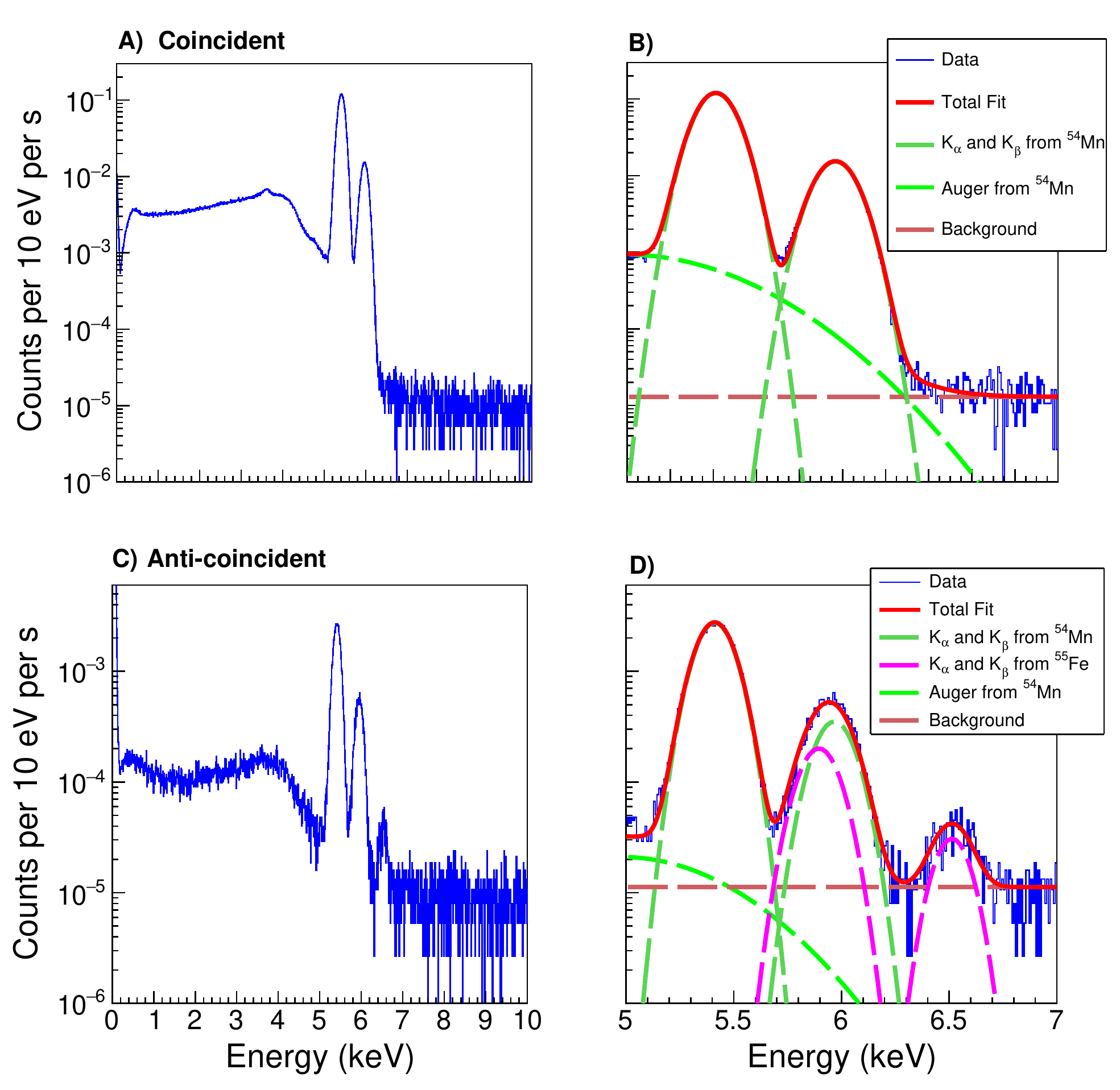}
    \centering
	\caption{\label{Fig:Efficiency_2us_Fit_Spec.eps} A) SDD energy spectrum for the coincident events from the \Mn\ source. The experimental data is from a 2~$\mu$s coincidence window. The \Kalpha~and \Kbeta~peaks are clearly visible at 5.41~keV and 5.96~keV as well as the silicon escape peak at 3.67~keV. B) Likelihood fit of the coincident data in the 5.0--7.0~keV region. C) SDD energy spectrum for the anti-coincident events from the \Mn\ source. The \Fe~contamination is visible at 6.51~keV. D) Likelihood fit of the anti-coincidence events in the 5.0--7.0~keV range.  See text for details of the fit components.}
\end{figure}

The gamma tagging efficiency ($\varepsilon_\gamma$) is defined as the probability that MTAS will detect the gamma when triggered by the X-ray (or Auger) in the SDD. \Mn~is an ideal source to study this parameter because it decays almost exclusively through the excited state, giving a $\rho$ (expected ratio of EC~to \ECStar~events) that is very small: $\rho= 3.0 \times 10^{-6}$. This value and the decay scheme of \Mn~are acquired from~\cite{be_table_2004}. The small ground state electron capture implies that there should be no false negatives (where an EC event looks like a \ECStar) during the exposure, which will simplify the analysis. False positives and negatives are further discussed in~\ref{sec:FalseNegative}. In addition, the energy of the released $\gamma$-ray from the excited state decay is 835~keV which is within a factor of two of the \K~gamma, the ultimate goal of the KDK experiment. Sec.~\ref{subsec:Geant_Simulations} details how the efficiency determined for 835~keV gammas can be extrapolated to 1460~keV, and Sec.~\ref{subsec:Dead Time Considerations} details live time considerations involved with the extrapolation. The \Mn~coincident plot can be seen in Fig.~\ref{Fig:Mn54_Coincident_Plot_2us.eps} and the individual SDD spectra are shown in Fig.~\ref{Fig:Efficiency_2us_Fit_Spec.eps} for a 2~$\mu$s coincidence window. 

In order to extract the efficiency for detecting in MTAS the \keV{835} gamma from the \Mn~source when the SDD triggers on the X-ray, a likelihood fit was performed with the efficiency as a shared parameter between the coincident and anti-coincident spectra. As there should be very few EC events due to the decay scheme of \Mn, the dominant source of events in the anti-coincidence spectrum would be from  \ECStar\ decays whose gamma has not been tagged (i.e. a false positive).  The likelihood also accounts for coincidences with the background which would otherwise cause the efficiency to be overestimated as the coincidence window increases. More information about these coincidences can be found in~\ref{sec:FalseNegative}.  Only the fully collected X-ray energy region between 5.0 and 7.0~keV was considered for this analysis with no energy cuts being made on the MTAS events. The shape of the coincidence spectrum was modelled with two gaussians (for the \Kalpha\ and \Kbeta\ X-ray peaks of the \Mn), a gaussian containing primarily an Auger contribution,  and a flat line representing any background contribution. The Auger gaussian was required to have the same shape for both the coincidence and anti-coincidence spectrum. To within statistical uncertainties, the result of this analysis does not depend on the precise shape of this contribution as long as it is treated consistently in all spectra. 

It is possible for the \ECStar\ decay to produce a conversion electron instead of a $\gamma$-ray and a respective term was included in the likelihood fit in order to account for this effect. %
The internal conversion coefficient for excited states is the ratio of internal conversion decays to decays via gamma ray emission. If the transition is greater than 1022~keV, internal pair formation is also possible.
For a \Mn\ K-shell excitation, the internal conversion coefficient is $2.49(4)\times10^{-4}$~\cite{kibedi_evaluation_2008,band_diracfock_2002}.

The anti-coincidence spectrum is modelled in an identical way except that two additional gaussians were used to account for the \Kalpha\ and \Kbeta\ of an \Fe\ contamination. Such contaminations are common in the production of \Mn\ \cite{be_table_2004} and would only be present in the anti-coincidence region as they have no excited state partner. The result of the fit for the \Mn\ source with a 2~$\mu$s coincidence window using all modules of MTAS can be seen in Fig.~\ref{Fig:Efficiency_2us_Fit_Spec.eps}. The efficiency at 835~keV was determined to be $0.9778(1)$ at 2~$\mu$s coincidence window.

The same analysis was performed for three separate coincidence windows (1.0, 2.0 and 4.0 $\mu$s), with the results shown in Table~\ref{tab:Modular_Efficiency}.

\begin{table}[ht]

\centering
\begin{tabular}{cccc}\hline
& 1 $\mu$s & 2 $\mu$s & 4 $\mu$s \\\hline
Total & 0.9775(1) & 0.9778(1) & 0.9778(1) \\\hline
\end{tabular}
\caption[Module MTAS Analysis]{\label{tab:Modular_Efficiency} Efficiency for detecting in MTAS the \keV{835} gamma from the \Mn~source when the SDD triggers on the X-ray, as a function of coincidence window.}
\end{table}

\subsection{\label{subsec:Geant_Simulations}Extrapolating the tagging efficiency to higher energies with Geant simulations}

In Section~\ref{subsubsec:Mn_Exp_Eff}, we measured the tagging efficiency for an 835~keV gamma over a selection of coincidence windows. However, for the KDK experiment the efficiency at 1460~keV is required. This is achieved by extrapolating the measured \Mn~efficiency with a \K/\Mn~efficiency ratio calculated using Geant4 (version 10.2p01)~\cite{collaboration2003geant4}.

 The SDD/MTAS setup was modelled in the software and a \Mn~and \K~source were individually simulated for ten million gamma events. The geometry of the source was exactly as described in Section~\ref{subsubsec:K40Source} and the initial position of the events was uniformly distributed with isotropic directional vectors. The number of gamma events from the source that deposit energy into any MTAS module above selected thresholds (anywhere from 0--50~keV) was recorded. The simulated efficiencies ($\varepsilon^s_{Mn}$ and $\varepsilon^s_{K}$) were determined by taking that number and dividing by the initial number of events. The simulated ratio between the \K~efficiency and \Mn\ efficiency ($\frac{\varepsilon^s_{K}}{\varepsilon^s_{Mn}}$) was then calculated. The procedure described above was repeated to extrapolate a tagging efficiency for \Zn\ (1115~keV), and the results of both analyses are shown in Table~\ref{tab:Geant_Total_Efficiency}. Simulated ratios were found to be very robust as they were independent (within one standard deviation) of energy threshold selections, different Geant4 physics lists (Penelope and Livermore) and large variations in material thickness surrounding the source.

As can be seen in Table~\ref{tab:Geant_Total_Efficiency}, the scaled efficiencies for both \Zn\ and \K\ are higher than that of the lower gamma energy \Mn\ source. This result, at the one part in a thousand correction, is due to the near 100$\%$ efficiency (MTAS radius $\sim$10 times the attenuation length of \K\ in NaI) and modularity of the MTAS detector.

\begin{table}[ht]
\centering
\resizebox{\linewidth}{!}{%
\begin{tabular}{ccccc}
\hline
\multirow{2}{*}{Isotope} &   Measured $\varepsilon^m_i$    &   \multirow{2}{*}{Simulated $\varepsilon^s_i$}   &   \multirow{2}{*}{Simulated Ratio $\frac{\varepsilon^s_i}{\varepsilon^s_{Mn}}$} &   Scaled $\frac{\varepsilon^s_i }{\varepsilon^s_{Mn}}\varepsilon^m_{Mn}$   \\
    &   (2~$\mu$s CW)  &   &  & (2~$\mu$s CW)  \\
\hline
\Mn     &   0.9778(1)   &   0.9847(3)   &   -   &   -   \\
\K      &   -   &   0.9860(3)   &   1.0013(5)   &   0.9791(5)   \\
\Zn     &   -   &   0.9866(3)   &   1.0019(5)   &   0.9797(5)   \\
\hline
\end{tabular}
}
\caption[]{\label{tab:Geant_Total_Efficiency} Extrapolation of MTAS gamma-tagging efficiencies from \Mn\ (835~keV) to \K\ (1460~keV) and \Zn\ (1115~keV) using Geant4 simulations. The experimental value is taken at 2~$\mu$s from Table~\ref{tab:Modular_Efficiency}. Due to the gammas typically arriving within less than 100~$ns$ the simulated ratios calculated above are valid for all coincidence windows, above that nominal threshold.}
\label{tab:extrap_efficiency}
\end{table}

\subsection{MTAS gamma spectrum components}\label{subsec:Geant_Components}

\begin{figure}[ht]
    \includegraphics[width=1.0\linewidth]{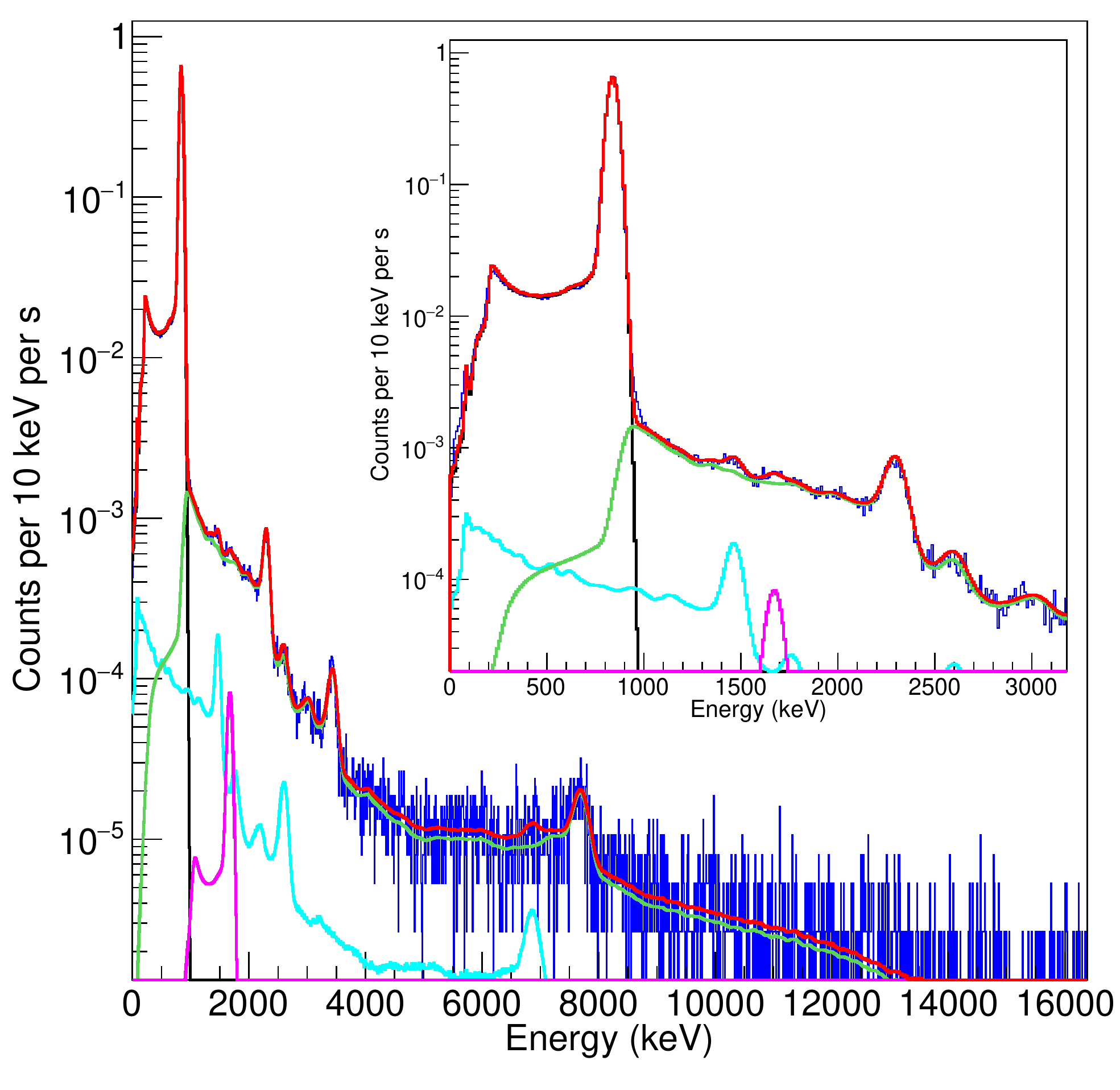}
    \centering
	\caption{\label{Fig:Mn54_Mn54_Gamma_Convolution.eps} Full \Mn,~4 $\mu$s coincidence window, gamma spectrum in MTAS when triggered by the SDD (dark blue) compared with Geant4 simulated data. All modules of MTAS are used. Components of the simulated spectrum include: The Geant4 single gamma deposition simulation (black), the MTAS natural background spectrum (light blue), the source-background convolution spectrum (green) and the source-source convolution spectrum (purple). These components are fit (total simulated spectrum in red) to the data to show their individual contributions to the experimental spectrum. Insert shows zoom of lower energy region.}
\end{figure}

The simulations allow us to study the different components that make up the SDD-triggered MTAS data spectrum, shown in Fig.~\ref{Fig:Mn54_Mn54_Gamma_Convolution.eps} for a \Mn~source. In this spectrum the single deposition of the source gamma is clearly identifiable by the photoelectric peak and Compton continuum. The tail ($>900$~keV) contains: i) random coincidences with only the natural MTAS background; ii) random coincidences convolving the MTAS background and the source gamma; and iii) random coincidences convolving two source gammas (eg. 2$\times$835 keV line). Fig.~\ref{Fig:Mn54_Mn54_Gamma_Convolution.eps} shows the different levels that each one of these components contribute to the overall spectrum. Other sources with more complicated decay schemes could include contributions from the $\beta^-$ or $\beta^+$ decay channels.

\subsection{Live time correction}\label{subsec:Dead Time Considerations}

Due to the level of accuracy required for this project, corrections for data losses need to be applied.
There are four physical pathway groupings in the digitizer: channel, channel group of four, Pixie module (16 channels), and the crate. It is important to note that there are no sources of dead-time for the channel group or for the Pixie Module. Our data rates in the crate were 2-3 orders of magnitude smaller than the level where this loss would be considerable. Furthermore, this would be a global loss rate and could not affect the ratio of EC/EC*. Channel level live times are the only factors to consider.

There are three sources of acquisition dead-time and two sources of processing dead-time to consider for this experiment. Acquisition dead times can be incurred in the digital signal processing (DSP) pathway, by the length of the trace, and from any shaping time. Acquisition dead-times only related to the SDD do not impact the EC/\ECStar\ ratio because both EC and EC* require an SDD trigger. This excludes the need to consider the trace and shaping related dead-times. The DSP live time can be calculated from the count rate. Data filtering from processing also affects our live time. Signal pile-up (when two signals are detected by the digitizer in the range of the slow trapezoidal filter) and event pile-up (when multiple signals are detected in the same channel and are considered in coincidence) are accounted for. We also consider the potential dead time associated with the multiplicity of events in the centre module of MTAS. 

All of these dead-times are modeled as the paralyzable type~\cite{xia_llc_pixie-16_2009}. This means that each occurrence of the situation leading to dead-time will prolong the length of the dead-time. This model means that the Output Count Rate (OCR) is related to the Input Count Rate ($ICR$) and the Dead Time ($dt$) by 

\begin{equation}\label{Eqn:OCR}
    OCR = ICR \cdot e^{-ICR \cdot dt } .
\end{equation}
This function is invertable with the use of the Lambert-W function as
\begin{equation}\label{Eqn:ICR}
ICR = W(-dt \cdot OCR) / -dt .
\end{equation}
The dead-time percentage can then be calculated as 
\begin{equation}\label{Eqn:DeadTime}
(1 - OCR/ICR) \cdot 100 .
\end{equation}
$OCR$ in the SDD and various rings of MTAS are shown in Table~\ref{tab:Rate_Table}. 

\begin{table*}[ht]
\centering
\begin{tabular}{ccccccc}\hline
Isotope & \multicolumn{6}{c}{Average Output Count Rates (OCR) (Hz)}\\
 & Plug & Central & Inner & Middle & Outer & SDD \\\hline
\Y & 225 & 1545 & 381 & 219 & 211 & 239   \\
\Zn & 27.6 & 309 & 174 & 198 & 204 & 46.5 \\
\Mn & 18.4 & 217 & 166 & 200 & 206 & 29.9 \\
\K &  12.7 & 160 & 159 & 196 & 202 & 11.9 \\\hline
\end{tabular}
\caption[Rate Table]{\label{tab:Rate_Table} Table of rates in various detector segments during measurements of different sources.  Errors are less than 0.01\%.}
\end{table*}

The live time correction factor for the total MTAS efficiency is calculated as a weighted average of the live time correction factors of the various rings of MTAS (Centre, Inner, Middle or Outer). The weight is taken to be the percentage that each ring contributes to the total efficiency, which is acquired from the simulations discussed in \ref{subsec:Geant_Simulations}. This yields the values listed in Table~\ref{tab:DT:live_times}. Live time correction is applied to the gamma-tagging efficiency (shown in Table~\ref{tab:Geant_Total_Efficiency}) via the ratio of the live time for the Mn isotope to that of the pertinent isotope (shown in Table~\ref{tab:DT:live_times}), which accounts for differing activities across the sources. The final energy and live time corrected efficiency value for the three coincidence windows are given in Table~\ref{tab:Final_Corrected_Efficiency}. This correction is of the order of 0.01$\%$.

\begin{table*}[ht]
\centering
\begin{tabular}{cccccc}
\hline
CW ($\mu$s)	&	\multicolumn{5}{c}{Total Live Times} \\ %\cline{2-6}
	&	\Mn	&	\K	&	\Zn	&	\Mn/\K	&	\Mn/\Zn	\\
\hline
1	&	0.9983(3)	&	 0.9982(4)	&	0.9987(3)	&	1.0001(4)	&	0.9996(3)	\\
2	&	0.9982(3)	&	 0.9980(4)	&	0.9986(3)	&	1.0001(4)	&	0.9996(3)	\\
4	&	0.9978(3)	&	 0.9977(4)	&	0.9982(3)	&	1.0001(4)	&	0.9996(3)   \\
\hline

\end{tabular}
\caption{Total live times for each isotope and correction factors Mn/K and Mn/Zn for various coincidence windows (CWs).}
\label{tab:DT:live_times}
\end{table*}

\begin{table}[ht]
\centering
\begin{tabular}{cccc}
\hline
CW ($\mu$s)	&	\multicolumn{3}{c}{Energy \& Live Time Corrected Efficiency} \\ %\cline{2-6}
	&	\Mn	&	\K	&	\Zn		\\
\hline
1	& 0.9775 (1) & \ \ 0.9789 (6)	&  0.9790 (6)		\\
2	& 0.9778 (1) & \ \ 0.9792 (6)	&	0.9793 (6)	\\
4	& 0.9778 (1) & \ \ 0.9792 (6)	&	0.9793 (6)	\\
\hline

\end{tabular}
\caption{Energy and live time corrected efficiency for \K~and \Zn~at three difference coincidence windows. Correction factors for energy and live time can be found in Table~\ref{tab:Geant_Total_Efficiency} and Table~\ref{tab:DT:live_times} respectively.}
\label{tab:Final_Corrected_Efficiency}
\end{table}

\subsection{Predicted sensitivity}\label{subsec:Predicted_Sensitivity}
The sensitivity of the composite method was studied using the blinded 44-day \K\ data set. To reduce biases, the data in the EC signal region (2.0 -- 3.8~keV in the anti-coincidence spectrum) and in the EC silicon escape peak (0.88 -- 1.4 keV) are hidden until the analysis is finalized. This data set was then used as a model to simulate the anti-coincidence and coincidence spectra over the 2--3.85~keV energy range. The blinding on the anti-coincidence signal region was maintained, forcing certain assumptions about the background. These include the linearity of the $\beta^-$ background, and  potential Cl and K fluorescence (distributed as Gaussians around the specific energy values). The signal (or $\rho$ value) is an input parameter of the simulation. The gamma tagging efficiency from Table~\ref{tab:Final_Corrected_Efficiency} and the probability of false negatives and positives from \ref{sec:FalseNegative} are included in the analysis. 

\begin{figure}[ht]
    \includegraphics[width=1.0\linewidth]{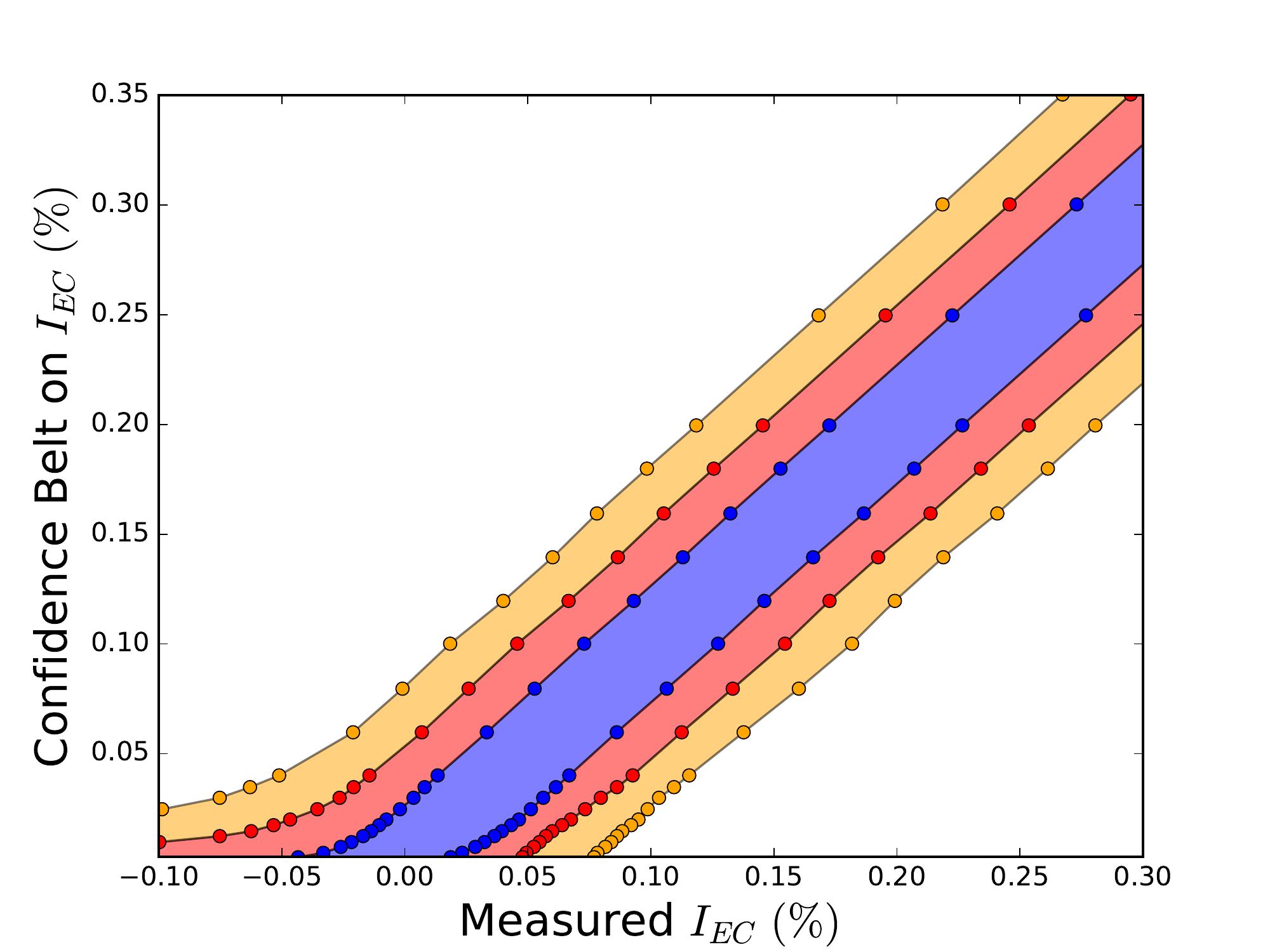}
    \centering
	\caption{\label{Fig:Feldman_And_Cousins_Confidence_Belt.png} Confidence belt of the statistical fluctuations of the simulated 44 day \K\ data set. The belt is constructed using the Feldman and Cousins ordering method. The 68.3, 95.4 and 99.7$\%$ confidence regions are shown in blue, red and orange respectively. The x and y axes are labelled by decay intensity, which is $\rho$\BRECStar. }
\end{figure}

By generating numerous simulations over a range of fixed $\rho$ input parameters, a frequentist confidence belt is constructed using the Feldman and Cousins ordering method~\cite{feldman1998unified}. This is shown in Fig.~\ref{Fig:Feldman_And_Cousins_Confidence_Belt.png}. This confidence belt is purely statistical and does not contain any systematic errors. In addition, a likelihood ratio analysis was performed on each simulation in order to determine a $p$-value when comparing the null hypothesis ($\rho = 0$ fixed, i.e. no decay to ground state) versus the alternative hypothesis ($\rho$ free). The $p$-value is defined as the probability of observing a test statistic (here, twice the logarithm of the  ratio of maximized likelihoods) at least as large as the one calculated assuming the null hypothesis is true.

The original design goal of the KDK experiment was to measure a branching ratio of 0.2$\%$~\cite{pradler_unverified_2013,be_table_2010}. Taking this value on the confidence belt in Fig.~\ref{Fig:Feldman_And_Cousins_Confidence_Belt.png} would generate a measurement of (0.2 $\pm~0.03) \%$ at a 68.3\% confidence level. In addition, with 0.2$\%$ as the input parameter we expect to reject the null hypothesis with a
$p$-value of $1.84 \times 10^{-13}$, corresponding to a $7.27\sigma$ significance in terms of upwards fluctuations of a gaussian.
%1-p=99.9999851\%$ confidence level or $5.1\sigma$ significance in terms of upwards fluctuations of a gaussian.
%
This result clearly indicates that the composite method can achieve the initial design goal of the KDK experiment. The full \K\ unblinding analysis will be published in a future paper. 

\section{\label{sec:Conclusion}Conclusion}

 This work has shown the design, commissioning and characterization of two experimental setups to measure the electron-capture decay of \K\ to its ground state. This measurement will be relevant to the fields of rare-event searches,  nuclear physics, and geochronological dating.
 
 For the composite method, the silicon drift detector (SDD) has been successfully constructed and integrated into the large Modular Total Absorption Spectrometer (MTAS). The SDD was calibrated with  radioactive isotopes (\Mn, \Zn, \Y\ and \K) over the 0.9--12~keV energy range. The energy threshold (370~eV) of the SDD was found to be well below the $\sim 3$~keV energy  of the \Ar\ K-shell X-rays and Auger electrons produced by the electron capture. The $\sim 200$~eV resolution of the detector will also resolve the \Kalpha\ and \Kbeta\ lines from the \K\ source. A \Mn\ source was used to determine the gamma ray tagging efficiency of the setup at 835~keV to a high level of precision. A robust simulation and live time analysis were performed to scale this efficiency to higher energies. This generated a final efficiency of 97.89(6)$\%$ for \K\ at a 1~$\mu$s coincidence window. 
The high value of the efficiency and its precision limit the impact of untagged \ECStar\ events as background.
Possible cases of mis-identified events (see~\ref{sec:FalseNegative}) were studied and characterized.
 
 A complementary, homogeneous method was also presented using a \KSI\ scintillator and dual PMT readout as the inner detector. The KSI crystal benefits from a large \K~content and almost total X-ray and Auger detection efficiency. The detector was calibrated using \Cs, internal \K, and Eu internal activation. Due to limited technical detail of the PMTs attached to the scintillator, work is being performed to replace them with better known SiPMs, with plans for a future measurement. 

With the composite method, a 44 day physics run has been completed using an enriched, thermally deposited, KCl source. The sensitivity will allow a successful branching ratio measurement if the true value is similar to the predicted 0.2$\%$. The isotopes \Zn\ and \Y\ were also measured for 1.4 and 0.5~days respectively, for future analysis. Our well-characterized setup should also allow the study of other isotopes.
%\appendix

\section{\label{sec:Acknowledgments}Acknowledgements}

Marek Karny provided thoughtful comments on an early version of this manuscript.
John Carter and Ryan Ickert provided stimulating input on the topic of geochronology. 
Engineering support has been contributed by Miles Constable and Fabrice R\'eti\`ere of TRIUMF, as well as by Koby Dering through the NSERC/Queen’s MRS. Paul Davis designed and supplied an earlier version of the electronics through the NSERC/University of Alberta MRS. 

Work was performed at Oak Ridge National Laboratory, managed by UT-Battelle, LLC, for the U.S. Department of Energy under Contract DE-AC05-00OR22725. Thermal deposition was conducted at the Center for Nanophase Materials Sciences, which is a DOE Office of Science User Facility. 
This manuscript has been authored by UT-Battelle, LLC under Contract No. DE-AC05-00OR22725 with the U.S. Department of Energy. The United States Government retains and the publisher, by accepting the article for publication, acknowledges that the United States Government retains a non-exclusive, paid-up, irrevocable, world-wide license to publish or reproduce the published form of this manuscript, or allow others to do so, for United States Government purposes. The Department of Energy will provide public access to these results of federally sponsored research in accordance with the DOE Public Access Plan (http://energy.gov/downloads/doe-public-access-plan).
Funding in Canada has been provided by NSERC through SAPIN and SAP RTI grants, as well as by the Faculty of Arts and Science of Queen's University, and by the McDonald Institute.
US support has also been supplied by the Joint Institute for Nuclear Physics and Applications.
This material is based upon work supported by the U.S. Department of Homeland Security under grant no. 2014-DN-077-ARI088-01. Disclaimer: The views and conclusions contained in this document are those of the authors and should not be interpreted as necessarily representing the official policies, either expressed or implied, of the U.S. Department of Homeland Security.

\appendix
\section{\label{sec:FalseNegative}Classifying events}

Our analysis involves determining the number of EC events relative to that of \ECStar~events, to obtain the ratio of the two branching ratios, $\rho = \frac{I_{EC}}{I_{EC^{*}}}$.
Experimentally, by fitting the spectra obtained in the SDD, we have access to the number of uncoincident and coincident events which ideally, would be proxies for the number of EC and \ECStar~events.
In reality, for various reasons, some EC events can be misidentified as \ECStar, and vice-versa.  In addition, both types of events can be lost completely while background events can look like either population.  We discuss these various events in the following text.

\subsection{\texorpdfstring{\ECStar\ misidentified as EC}{EC* misidentified as EC because of missed gamma}}

For a source of activity $A$, and branching ratio \BRECStar, the expected rate of SDD triggers is $A$ \BRECStar $\eta$, where $\eta$ is the fraction of X-rays/Augers escaping the source, reaching the SDD, and triggering it.
Because the tagging efficiency ($\varepsilon$) of MTAS for $\gamma$s and conversion electrons from the excited state is not quite 100\%, an \ECStar\ can look like an EC.  
Since $\gamma$-rays dominate the relaxation, 
the expected rate of such events is $A$ \BRECStar $\eta$$(1-\varepsilon_{\gamma})$, while the rate of properly identified EC* events is $A$ \BRECStar $\eta$$\varepsilon_\gamma$.
For our $\sim98$\% efficiencies (Sec.~\ref{subsec:Dead Time Considerations}), assuming $\rho=0.02$ leads to as many events from this background as from the signal.

\subsection{EC misidentified as EC* because of coincidences with MTAS background}\label{App:BmT}
Spurious coincidences between EC triggers in the SDD and events in MTAS can cause the EC to be misidentified as \ECStar.  Since the rate of MTAS background \BMTAS$ \approx 2.4$~kHz (Sec.~\ref{subsec:MTAS}) is higher than the activity of the sources used (Table~\ref{tab:SDD_Energy_Calibration}), the MTAS background is the dominant factor.

\begin{figure}[ht]
    \includegraphics[width=1.0\linewidth]{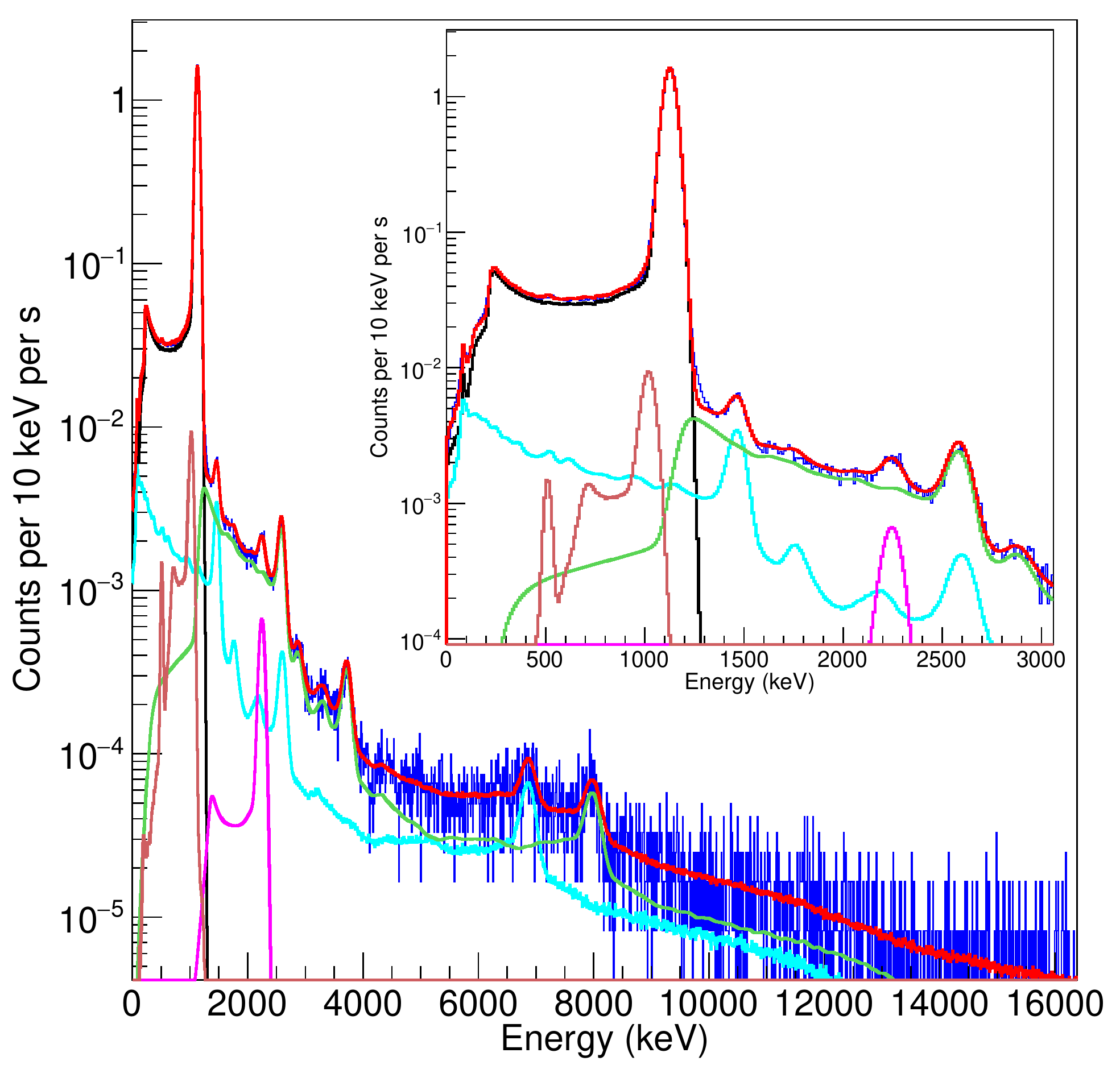}
    \centering
	\caption{\label{Fig:MTAS_Zn65_False_Negative.eps} Full \Zn,~4 $\mu$s coincidence window, gamma spectrum in MTAS when triggered by the SDD (dark blue) compared with Geant4 simulated data. All modules of MTAS are used. Components of the simulated spectrum include: The Geant4 single gamma deposition simulation (black, 0-background coincidence spectrum), the MTAS natural background spectrum (light blue), the source-background convolution spectrum (green, 1-background coincidence spectrum), the source-source convolution spectrum (purple) and the $\beta+$ annihilation spectrum (brown). These components are fit (total simulated spectrum in red) to the data to show their individual contributions to the experimental spectrum. Insert shows zoom of lower energy region. }
\end{figure}

The expected number of background events in an arbitrary time window $T \leq 4 \ \mu$s is given by \BMTAS $T \lesssim 0.01 \ll 1$. 
In practice, though the rate of background \BMTAS \ is well known, the effective coincidence window, $T$, may differ from the nominal value due to the rolling coincidence window. We can, however, use coincidences between calibration data and background to determine the product \BMTAS $T$. 
The Poisson probability of having no spurious coincidences after an EC SDD trigger is  $P_0 = e^{-B_{M} T}$, and the probability of having exactly one spurious coincidence is $P_1 = e^{-B_{M} T}B_{M} T$.   If the rate of SDD events from the source is $\sigma$, then the rate of source events with 0 background coincidence is $\sigma P_0$, and the rate of source events with  1 background coincidence is $\sigma P_1$.  The ratio of these two populations is $P_1/P_0 = B_{M} T$.  This number can be determined by considering the SDD triggered MTAS spectrum.  
Neglecting higher-order source-background coincidences, the spectrum is the sum of the 0-coincidence spectrum and of the 1-coincidence one, the latter being the convolution of the 0-coincidence spectrum and of the background spectrum.  This is illustrated in Fig.~\ref{Fig:Mn54_Mn54_Gamma_Convolution.eps} for \Mn,  and Fig.~\ref{Fig:MTAS_Zn65_False_Negative.eps} for \Zn. 
The results are listed in Table~\ref{tab:Beta_T_Tab}, which shows a strong agreement between the \BMTAS$T$ values for both \Mn~and~\Zn, as expected from this parameter that should be independent of the source isotope and activity.

\begin{table*}[ht]
\centering
\begin{tabular}{ccccc}
\hline
Nominal Coincidence Window ($\mu$s) & $^{65}$Zn \BMTAS$T$ & $^{54}$Mn \BMTAS$T$ & \BSDD \BMTAS $T$ (Hz) & \BSDD (1-\BMTAS $T$) (Hz) \\\hline
1.0  & 0.0075(1) & 0.0074(1) & 4.0(6)$\times 10^{-6}$ & 5.4(8)$\times 10^{-4}$\\
2.0  & 0.0126(1) & 0.0125(1) & 6.8(10)$\times 10^{-6}$ & 5.3(8)$\times 10^{-4}$\\
4.0  & 0.0230(2) & 0.0227(2) & 1.2(2)$\times 10^{-5}$ & 5.3(8)$\times 10^{-4}$\\\hline

\end{tabular}
\caption[]{\label{tab:Beta_T_Tab} Measured  \BMTAS$T$ values for the different sources as a function of coincidence windows. Also shown are the expected rate of SDD background events that look like \ECStar (\BSDD \BMTAS $T$) and EC (\BSDD (1-\BMTAS$T$)) events  as a function of coincidence window. The values of \BMTAS$T$ are taken from the \Mn\ data. }
\end{table*}

\subsection{SDD background}
The SDD background, illustrated in Fig.~\ref{Fig:Zn65_vs_BKG_Spectrum}, has a rate of \BSDD $=5.4(8)~\times~10^{-4} $~Hz in the  2-6 keV energy region. The expected rate of events that are not in coincidence with gamma background, and that can therefore look like EC events, is  \BSDD$(1-$\BMTAS $T)$.  The expected rate of events that can look like an EC* because they are in coincidence with the gamma background is \BSDD\BMTAS$T$.  Values are provided in Table~\ref{tab:Beta_T_Tab}, taking the values of \BMTAS T from \Mn\ in the same table.

\subsection{\label{subsec:SDD_Efficiency}Source tagging efficiency}

\begin{figure}[ht]
    \includegraphics[width=0.9\linewidth]{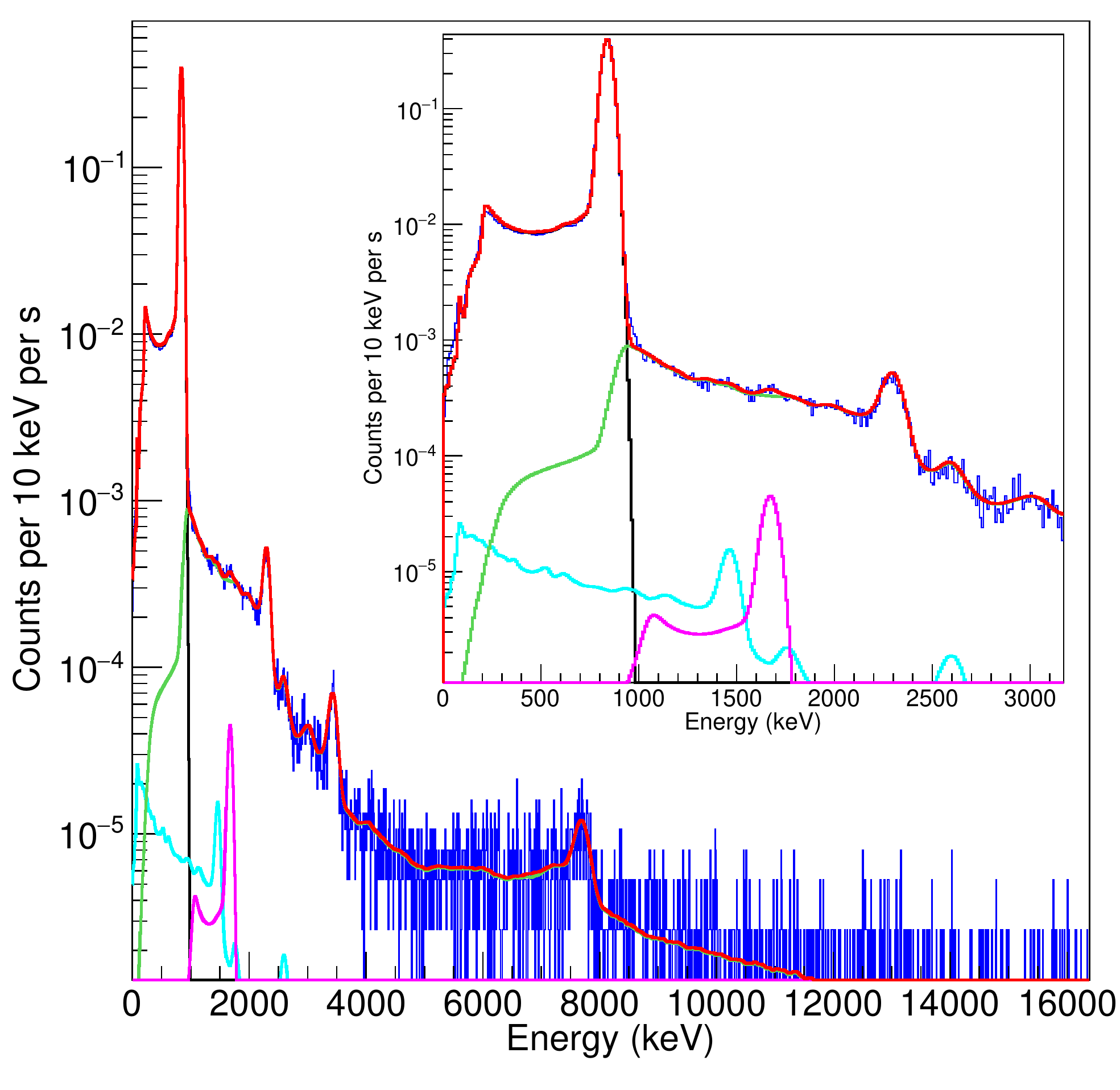}
    
	\caption{\label{Fig:X_Ray_Cut_Mn54_Mn54_Gamma_Convolution.eps} \Mn,~4 $\mu$s coincidence window, gamma spectrum in MTAS when triggered by the SDD (dark blue) with an energy gate of 5.0 - 7.0~keV. All modules of MTAS are used. Components of the simulated spectrum include: The Geant4 single gamma deposition simulation (black), the MTAS natural background spectrum (light blue), the source-background convolution spectrum (green) and the source-source convolution spectrum (purple). These components are fit (total simulated spectrum in red) to the data to show their individual contributions to the experimental spectrum. Insert shows zoom of lower energy region.}
\end{figure}

\begin{figure}[ht]
    \includegraphics[width=0.9\linewidth]{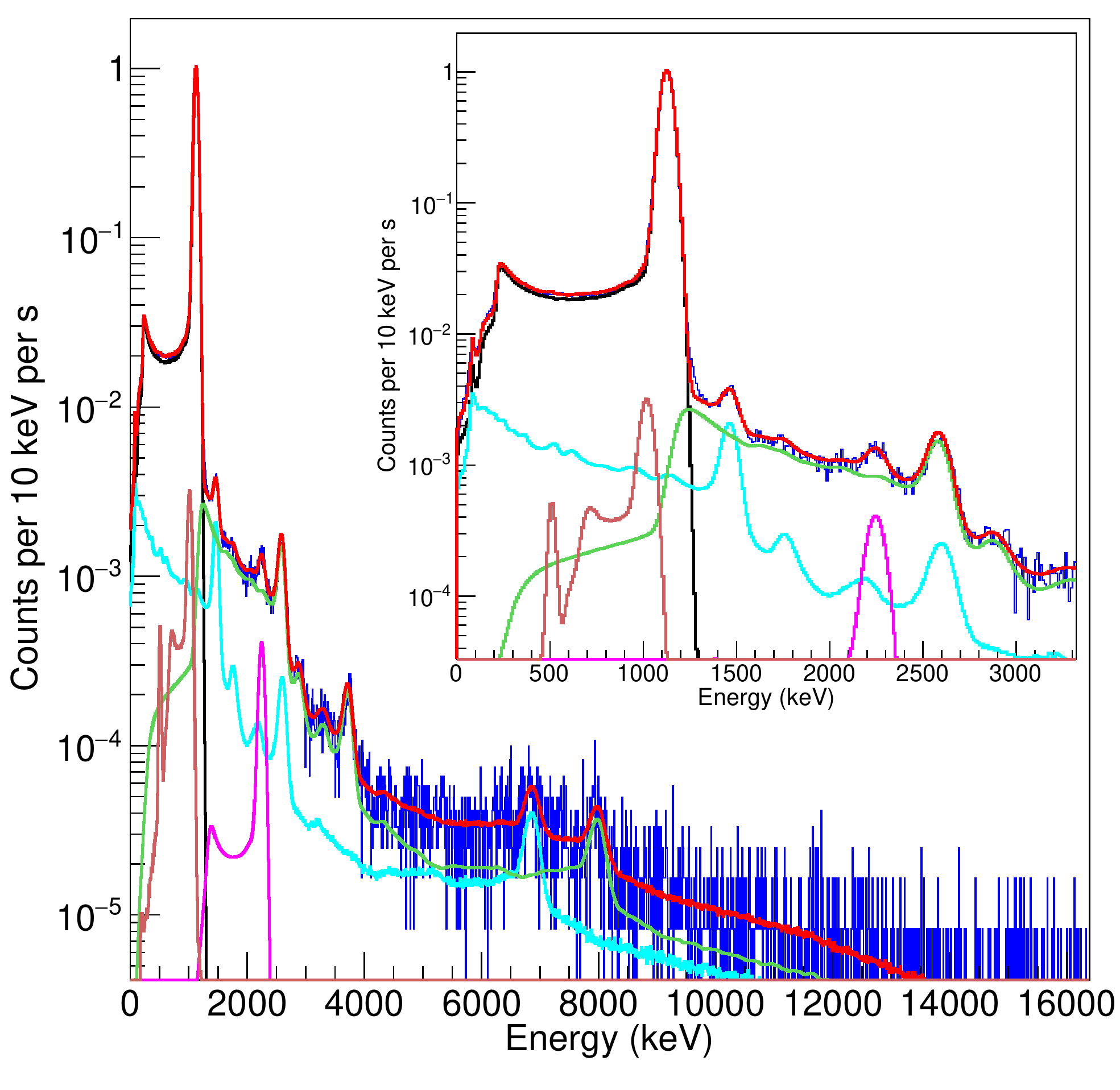}
    \centering
	\caption{\label{Fig:X_Ray_Cut_MTAS_Zn65_False_Negative.eps} \Zn,~4 $\mu$s coincidence window, gamma spectrum in MTAS when triggered by the SDD (dark blue) with an energy gate of 7.7 - 9.5 keV. All modules of MTAS are used. Components of the simulated spectrum include: The Geant4 single gamma deposition simulation (black), the MTAS natural background spectrum (light blue), the source-background convolution spectrum (green), the source-source convolution spectrum (purple) and the $\beta+$ annihilation spectrum (brown). These components are fit (total simulated spectrum in red) to the data to show their individual contributions to the experimental spectrum. Insert shows zoom of lower energy region. }
\end{figure}

It is possible for an EC event to be detected in the SDD in spurious coincidence with a gamma from an \ECStar~whose X-ray/Auger was missed. In this case the EC event would be incorrectly classified as an \ECStar. The expected rate of this occurring is dependent on the X-ray/Auger tagging efficiency ($\eta$). This efficiency is the probability that when an electron capture decay occurs, the X-ray/Auger will make it out of the source, into the solid angle of the SDD, past the dead layer of the SDD and deposit some energy into the detector. If an energy selection is imposed on the X-rays/Augers that arrive on the SDD, we label this efficiency $\eta_E$. Although simulations can provide a value for $\eta_E$, an experimental measurement can be made by looking at the SDD-triggered MTAS spectrum with X-ray/Auger selection cuts (seen in Fig.~\ref{Fig:X_Ray_Cut_Mn54_Mn54_Gamma_Convolution.eps} and Fig.~\ref{Fig:X_Ray_Cut_MTAS_Zn65_False_Negative.eps}). The area ($x$, counts/s) under the single gamma (black) spectrum in Fig.~\ref{Fig:X_Ray_Cut_Mn54_Mn54_Gamma_Convolution.eps} and Fig.~\ref{Fig:X_Ray_Cut_MTAS_Zn65_False_Negative.eps} is given as the activity ($A$) of the source multiplied by branching ratio and the probability of detecting the X-ray/Auger and the gamma (\BRECStar~$\varepsilon_\gamma~\eta$):  
\begin{equation}\label{Eqn:MTAS_Single_Gamma_Spectrum}
    x = A~I_{EC^*} \ \varepsilon_\gamma \ \eta.
\end{equation}
With an energy selection on the SDD keeping mainly X-rays, this becomes
\begin{equation}\label{Eqn:MTAS_Single_Gamma_Spectrum_1}
    x = A~I_{EC^*} \ \varepsilon_\gamma \ \eta_E.
\end{equation}
In the same figures, the area ($y$, counts/s) under the purple curve represents events when two \ECStar~decays occur and either one or two X-rays from the two decays are detected, and both gammas are seen in MTAS. The area can be expressed as
\begin{equation}\label{Eqn:MTAS_Conv_Gamma_Spectrum_1}
    y = (A~I_{EC^*} \varepsilon_\gamma)^2 \  T  \ \left[ 
    \underbrace{2\eta(1-\eta)}_{\text{1~X-ray}}
    + 
    \underbrace{\eta^2}_{\text{both~X-rays}}
    \right],
\end{equation}
where $T$ is the true coincidence window. Due to the shape of the SDD spectrum,  if the  selection $\eta_E$ is made on the K-shell lines, the probability that two X-rays sum to this selection is negligible, leaving only the term coming from single interactions:
\begin{equation}\label{Eqn:MTAS_Conv_Gamma_Spectrum}
    y = (A~I_{EC^*} \varepsilon_\gamma)^2  T \ 2 \eta_E(1-\eta_E).
\end{equation}
Rearranging Eqn.~\ref{Eqn:MTAS_Single_Gamma_Spectrum_1} and Eqn.~\ref{Eqn:MTAS_Conv_Gamma_Spectrum}  solves for $\eta_E$:

\begin{equation}\label{Eqn:MTAS_Eta_x}
    \eta_E = \frac{2Tx^2}{(y+2Tx^2)}.
\end{equation}

The parameters $x$ and $y$ can be measured from the fit in Fig.~\ref{Fig:X_Ray_Cut_Mn54_Mn54_Gamma_Convolution.eps} and Fig.~\ref{Fig:X_Ray_Cut_MTAS_Zn65_False_Negative.eps}. The true coincidence window value (T) can be determined from \BMTAS, Table~\ref{tab:MTAS_BKG_Rate} and, \BMTAS$T$, Table~\ref{tab:Beta_T_Tab}. The measured $\eta_E$ values for \Mn~and \Zn~are shown in Table~\ref{tab:Eta_x}. A tight x-ray gate of 5.0 - 7.0~keV for the \Mn~and 7.7 - 9.5~keV for the \Zn~was chosen for this analysis.

\begin{table*}[ht]
\centering
\begin{tabular}{ccc}\hline
Nominal Coincidence Window ($\mu$s) & \Mn~$\eta_E$   & \Zn~$\eta_E$ \\\hline
1.0  & 31.1 $\pm$ 13.6 $\%$ & 20.2 $\pm$ 2.3 $\%$ \\
2.0  & 26.0 $\pm$ 7.5 $\%$ & 22.2 $\pm$ 2.1 $\%$ \\
4.0  & 24.0 $\pm$ 4.8 $\%$  & 21.8 $\pm$ 1.5 $\%$ \\
Weighted Average & 25.9 $\pm$ 3.9\% & 21.5 $\pm$ 1.1 \% \\\hline

\end{tabular}
\caption[]{\label{tab:Eta_x} Measured $\eta_E$ values for the \Zn~and \Mn~sources as a function of coincidence windows. The values for $\eta_E$ agree within one standard deviation as a function of coincidence window.}
\end{table*}

The values for $\eta_E$ shown in Table~\ref{tab:Eta_x} agree within one standard deviation when comparing different coincidence windows of the same source. The error on the $\eta_E$ calculated for \Mn~is high due to the small integral of the source-source convolution component (see Fig.~\ref{Fig:X_Ray_Cut_Mn54_Mn54_Gamma_Convolution.eps}). This is due to the lower activity of the \Mn~source. Additionally, it is expected that the \Zn~source will have a higher efficiency than the \Mn~source due to its higher X-ray energy being more likely to penetrate past the dead layer of the detector. However, this measurement is also dependent on the source geometry and construction which could be different between our isotopes. 

Other effects, due to coincidences with the various radioactive sources and their decay channels, are possible, and will be addressed in publications pertaining to those sources.

%\subfile{Subfiles/M54_Efficiency_Model.tex}
%\subfile{Subfiles/Appendix_KCl_Calculation}
%\subfile{Subfiles/Extra_Parts.tex}

%\subfile{Subfiles/Appendix_For_live_time.tex}

%\subfile{Subfiles/Appendix_For_Geant.tex}
%\subfile{Subfiles/Appendix_Conversion_Electrons}

%\subfile{Subfiles/Appendix_KSI}

\bibliographystyle{elsarticle-num}
\bibliography{KDK_Bib_Edit.bib}

\end{document}